\documentclass{elsart}
\usepackage{algorithmic}
\usepackage{algorithm}

\usepackage{graphicx}
\usepackage{amssymb}
\usepackage{amsmath}
\usepackage{enumerate}
\usepackage{multirow} 
\usepackage{booktabs}
\usepackage[switch,pagewise]{lineno}
\usepackage[usenames]{color} 
\usepackage{url}
\usepackage{ulem}
\usepackage{rotating} 

\usepackage{threeparttable}

\begin{document}
\begin{frontmatter}

\title{An iterated local search algorithm for the minimum differential dispersion problem}

\author[Angers]{Yangming Zhou} and
\ead{zhou.yangming@yahoo.com}
\author[Angers,Paris]{Jin-Kao Hao\corauthref{cor}}
\corauth[cor]{Corresponding author.}
\ead{hao@info.univ-angers.fr}
\address[Angers]{LERIA, Universit{\'e} d'Angers, 2 Boulevard Lavoisier, 49045 Angers, France}
\address[Paris]{Institut Universitaire de France, Paris, France}
\maketitle

\begin{abstract}

Given a set of $n$ elements separated by a pairwise distance matrix, the minimum differential dispersion problem (\textsl{Min-Diff DP}) aims to identify a subset of $m$ elements ($m < n$) such that the difference between the maximum sum and the minimum sum of the inter-element distances between any two chosen elements is minimized. We propose an effective iterated local search (denoted by ILS\_MinDiff) for \textsl{Min-Diff DP}. To ensure an effective exploration and exploitation of the search space, the proposed ILS\_MinDiff algorithm iterates through three sequential search phases: a fast descent-based neighborhood search phase to find a local optimum from a given starting solution, a local optima exploring phase to visit nearby high-quality solutions around a given local optimum, and a local optima escaping phase to move away from the current search region. Experimental results on six data sets of 190 benchmark instances demonstrate that ILS\_MinDiff competes favorably with the state-of-the-art algorithms by finding 130 improved best results (new upper bounds). 
\normalem

\emph{keywords}: Combinatorial optimization; dispersion problems; heuristics; iterated local search; three phase search.
\end{abstract}
\end{frontmatter}

\section{Introduction}
\label{Sec:Introduction}


Let $N = \{e_1 ,e_2 ,\ldots,e_n\}$ be a set of $n$ elements and $d_{ij}$ be the distance between $e_i$ and $e_j$ according to a given distance metric such that $d_{ij} > 0$ if $ i \neq j$ and $d_{ij} = 0$ otherwise. 
The minimum differential dispersion problem (\textsl{Min-Diff DP}) is to identify a subset $S \subset N$ of a given cardinality $m$ $(m < n)$, such that the difference between the maximum sum and the minimum sum of the inter-element distances between any two elements in $S$ is minimized. Formally, \textsl{Min-Diff DP} can be described in the following way. 

Let $\Delta(e_v)$ be the sum of pairwise distances between an element $e_v \in S$ and the remaining elements in $S$, that is:

\begin{equation}
\Delta(e_v) = \sum_{e_u \in S, u \neq v} d_{uv}
\end{equation} 

The objective value $f$ of the solution $S$ is then defined by the following differential dispersion:

\begin{equation}
f(S) = \max_{e_u \in S} \{\Delta(e_u)\} - \min_{e_v \in S} \{\Delta(e_v)\}
\end{equation}

Then, \textsl{Min-Diff DP} is to find a subset $S^* \subset N$ of size $m$ with the minimum differential dispersion, i.e.,

\begin{equation}
S^* = \arg \min_{S \in \Omega} f(S)
\end{equation}

where $\Omega$ is the search space including all possible subsets of size $m$ in $N$, i.e., $\Omega = \{S: S \subset N~\text{and}~|S| = m\}$. The size of $\Omega$ is extremely large, up to a maximum number of $\binom{n}{m} = \frac{n!}{m!(n-m)!}$. 

\textsl{Min-Diff DP} is one of many diversity or dispersion problems \cite{Prokopyev2009} which basically aim to find a subset $S$ from a given set of elements, such that a distance-based objective function over the elements in $S$ is maximized or minimized. These problems can be further classified according to two types of objective functions:

\begin{itemize}
\item \textit{Efficiency-based} measures which consider some dispersion quantity for all elements in $S$. This category mainly includes the \textsl{maximum diversity problem} (MDP) and the \textsl{max-min diversity problem} (MMDP), which respectively maximizes the total sum of the inter-element distances of any two chosen elements and the minimum distance of any two chosen elements.
\item \textit{Equity-based} measures which guarantee equitable dispersion among the selected elements. This category includes three problems: (i) the \textsl{maximum mean dispersion problem} (Max-Mean DP) maximizes the average inter-element distance among the chosen elements; (ii) the \textsl{maximum min-sum dispersion problem} (Max-Min-sum DP) maximizes the minimum sum of the inter-element distances between any two chosen elements; (iii) the \textsl{minimum differential dispersion problem} considered in this work. It is worth noting that the cardinality of subset $S$ is fixed except for Max-Mean DP.
\end{itemize}

In addition to their theoretical significance as NP-hard problems, diversity problems have a variety of real-world applications in facility location \cite{Kuby1987}, pollution control \cite{Carrasco2015}, maximally diverse/similar group selection (e.g., biological diversity, admission policy formulation, committee formation, curriculum design, market planning) \cite{Ghosh1996,Marti2013,Mladenovic2016}, densest subgraph identification \cite{Kortsarz1993}, selection of homogeneous groups \cite{Brown1979b}, web pages ranking \cite{Kerchove2008,Yang2015}, community mining \cite{Yang2007}, and network flow problems \cite{Brown1979a}.

In this study, we focus on \textsl{Min-Diff DP}, which is known to be strongly NP-hard \cite{Prokopyev2009}. \textsl{Min-Diff DP} can be formulated as a 0-1 mixed integer programming problem (MIP) \cite{Prokopyev2009}. Thus it can be conveniently solved by MIP  solvers like IBM ILOG CPLEX Optimizer (CPLEX). However, being an exact solver, CPLEX is only able to solve instances of small size (up to $n = 40$ and $m = 15$), while requiring high CPU times (more than 2500 seconds) \cite{Prokopyev2009}. For medium and large instances, heuristic and meta-heuristic algorithms are often preferred to solve the problem approximately. In recent years, several heuristic approaches have been proposed in the literature \cite{Prokopyev2009,Duarte2015,Mladenovic2016}. In particular, in 2015, based on greedy randomized adaptive search procedure (GRASP), variable neighborhood search (VNS) and exterior path relinking (EPR), Duarte et al. proposed several effective hybrid heuristics \cite{Duarte2015}. Very recently (2016), Mladenovi{\'c} et al. proposed an improved VNS algorithm which uses the swap neighborhood both in its descent and shaking phases \cite{Mladenovic2016}. This new VNS algorithm significantly outperforms the previous best heuristics reported in \cite{Duarte2015} and is the current best-performing algorithm available in the literature for \textsl{Min-Diff DP}. We will use it as our main reference for the computational studies.


Our literature review showed that, contrary to other diversity problems like MDP and Max-Min DP for which many methods, both exact and heuristic, have been investigated, there are currently only a few studies for \textsl{Min-Diff DP}, in particular in terms of heuristic methods. To fill the gap, we introduce in this work an  iterated local search algorithm, denoted as ILS\_MinDiff, which adopts the general framework of the three-phase search. To efficiently explore the search space, ILS\_MinDiff iterates through three sequential search phases: a descent-based neighborhood search phase to reach a local optimum from a given starting solution, a local optima exploring phase to discover nearby local optima within a given search region and a local optima escaping phase to displace the search into a new and distant region. Despite its simplicity, ILS\_MinDiff competes very favorably with the state-of-the-art methods when it was tested on 190 benchmark instances available in the literature. Specifically, ILS\_MinDiff achieved improved best results (new lower bounds) for 131 out of 190 instances ($\approx 69\%$) and matched the best-known results for 42 instances.

The rest of the paper is organized as follows. In the next section, we present a brief literature review on the iterated local search framework and its two recent variants. In Section \ref{Sec:Hybrid Iterated Local Search}, we describe the general framework and the key components of the proposed algorithm. In Section \ref{Computational Experiments}, we present an extensive experimental comparison with state-of-the-art algorithms. A parameter analysis is provided in Section \ref{Sec:Experimental Analysis}, followed by conclusions in Section \ref{Sec:Conclusions}.

\section{Related work on applications of iterated local search}
\label{Sec:Related Work on Iterated Local Search}

\begin{figure}[!htbp]
\centering
\includegraphics[width=0.9\textwidth]{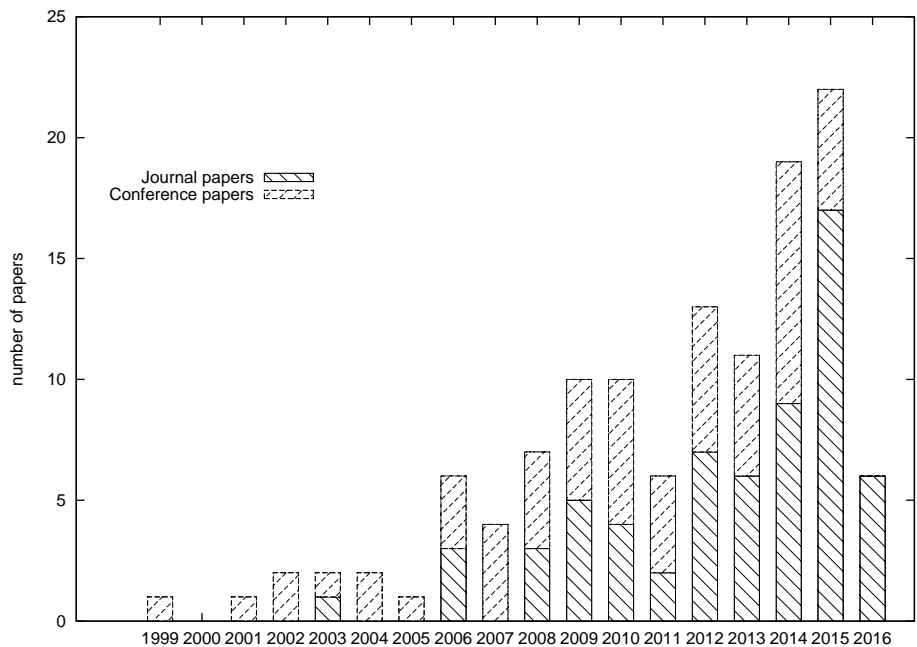}
\caption{Journal and conference publications by years on ILS. Data was extracted from the DBLP database http://dblp.uni-trier.de/search/publ under the keywords ``iterated local search" on 29 June 2016. }
\label{Fig:Paper Statistic}
\end{figure}

As one of the most widely-used meta-heuristic approaches, \textsl{Iterated local search} (ILS) \cite{Lourencco2003} has been successfully applied to solve a variety of combinatorial optimization problems. In spite of its conceptual simplicity, it has led to a number of state-of-the-art results. Figure  \ref{Fig:Paper Statistic} shows that over the last two decades, there has been an increasing interest in ILS, as witnessed by the number of publications related to ILS. 

ILS is an iterated two-phase approach whose key idea is to explore at each iteration the search zones around the last local optimum discovered by a local search procedure. Typically, ILS iteratively alters between two phases: a perturbation phase to modify the current local optimal solution followed by a local search phase to find a new local optimum from the modified solution (Algorithm \ref{Algorithm:ILS framework}) \cite{Lourencco2003}. 
 
\begin{algorithm}
\begin{small}
 \caption{Iterated local search}
 \label{Algorithm:ILS framework}
 \begin{algorithmic}[1]
 	\STATE $S_0 \leftarrow$  GenerateInitialSolution$()$\\
 	\STATE $S^* \leftarrow $ LocalSearch$(S_0)$ \\ 
	\WHILE{a stopping condition is not reached}
		\STATE $S' \leftarrow $ Perturbation$(S^*,history)$\\
		\STATE ${S^*}' \leftarrow $ LocalSearch$(S')$ \\ 
		\STATE $S^* \leftarrow $ AcceptanceCriterion$(S^*,{S^*}', history)$ \\ 
	\ENDWHILE
 \end{algorithmic}
 \end{small}
 \end{algorithm}

Based on the general ILS framework, several variants and extended approaches have recently been proposed in the literature, of which two representative examples are breakout local search (BLS) \cite{Benlic2012,Benlic2013d} and three-phase search (TPS) \cite{Fu2015}. The effectiveness of BLS and TPS have been verified on a variety of hard optimization problems and applications (see examples of Table \ref{Tab:Literature Statistic On BLS and TPS}). In the following, we present a brief review of these ILS variants.


\begin{table*}[!ht]
\caption{A summary of the applications of breakout local search and three-phase search.}
\label{Tab:Literature Statistic On BLS and TPS}
\vskip 0.15in
\begin{center}
\begin{scriptsize}
\begin{tabular}{l|l}
\toprule[0.75pt]
Breakout local search & Three-phase search\\
\midrule[0.5pt]
minimum sum coloring problem \cite{Benlic2012}  & quadratic minimum spanning tree problem \cite{Fu2015} \\
quadratic assignment problem \cite{Benlic2013a}  & maximally diverse grouping problem \cite{Lai2016} \\
maximum clique problem \cite{Benlic2013b} &  capacitated clustering problem \cite{Lai2016b} \\ 
max-cut problem \cite{Benlic2013c} & max-k-cut problem \cite{Ma2016} \\
vertex separator problem \cite{Benlic2013d} &  clique partitioning problem \cite{Zhou2015}\\
Steiner tree problem \cite{Fu2014}  &  minimum differential dispersion problem \\
assembly sequence planning problem \cite{Ghandi2015} & \\
single-machine total weighted tardiness problem \cite{Dingetal2016} & \\
\bottomrule[0.75pt]
\end{tabular}
\end{scriptsize}
\end{center}
\vskip -0.1in
\end{table*}

\textsl{Breakout local search} introduced in \cite{Benlic2012,Benlic2013d}  combines local search with a \textit{dedicated  and adaptive perturbation} mechanism. Its basic idea is to use a descent-based local search procedure to intensify the search in a given search region, and to perform  dedicated perturbations to jump into a new promising search region once a local optimum is encountered. BLS is characterized by its adaptive perturbation. At the perturbation phase, BLS attempts to achieve the most suitable degree of diversification by dynamically determining the number of perturbation moves (i.e., the jump magnitude) and by adaptively selecting between several types of pre-defined perturbation operations of different intensities, which is achieved through the use of information from specific memory structures. As summarized in Table \ref{Tab:Literature Statistic On BLS and TPS}, BLS has reported excellent performances for several well-known combinatorial optimization problems.  Algorithm \ref{Algorithm:BLS framework} describes the general framework of BLS. BLS distinguishes itself from the conventional ILS approach by the following two aspects. First, multiple types of perturbations are used in BLS, which are triggered according to the search states, achieving variable levels of diversification. Second, the local optimal solution returned by the local search procedure is always accepted as the new starting solution in BLS regardless of its quality, which completely eliminates the acceptance criterion component of ILS (Alg. \ref{Algorithm:ILS framework}, line 6).
 
\begin{algorithm}
\begin{small}
 \caption{Breakout local search}
 \label{Algorithm:BLS framework}
 \begin{algorithmic}[1]
 	\STATE $S \leftarrow$  GenerateInitialSolution$()$ \\
 	\STATE $L \leftarrow L_0$ \\ 
	\WHILE{a stopping condition is not reached}
		\STATE $S' \leftarrow $ DescentBasedSearch$(S)$ \\
		\STATE $L \leftarrow $ DetermineJumpMagnitude$(L,S',history)$ \\ 
		\STATE $T \leftarrow $ DeterminePerturbationType$(S',history)$ \\
		\STATE $S \leftarrow $ Perturb$(L,T,S', history)$ \\ 
	\ENDWHILE
 \end{algorithmic}
 \end{small}
 \end{algorithm}
 

\textsl{Three-phase search} proposed in \cite{Fu2015} follows and generalizes the basic ILS scheme. TPS iterates through three distinctive and sequential search phases. The basic idea of TPS is described as follows. Starting from an initial solution, a descent-based neighborhood search procedure is first employed to find a local optimal solution. Then, a local optima exploring phase is triggered with the purpose of discovering nearby local optima of better quality. When the search stagnates in the current search zone, TPS turns into a diversified perturbation phase, which strongly modifies the current solution to jump into a new search region. The process iteratively runs the above three phases until a given stopping condition is met. Compared to BLS, TPS further divides the perturbation phase into a local optima exploring phase (to discover more local optima within a given region) and a diversified perturbation phase (to displace the search to a new and distant search region). TPS has been successfully used to solve several optimization problems, as shown in the right column of Table \ref{Tab:Literature Statistic On BLS and TPS}. The general framework of TPS is outlined in Algorithm \ref{Algorithm:TPS framework}. Actually, the ILS\_MinDiff algorithm proposed in this work follows the TPS framework. 

\begin{algorithm}
\begin{small}
 \caption{Three-phase search}
 \label{Algorithm:TPS framework}
 \begin{algorithmic}[1]
 	\STATE $S \leftarrow$  GenerateInitialSolution$()$\\
 	\STATE $S^* \leftarrow S$\\ 
	\WHILE{a stopping condition is not reached}
		\STATE $S \leftarrow $ DescentBasedSearch$(S)$\\
		\STATE $S \leftarrow $ LocalOptimaExploring$(S,history)$ \\
		\STATE $S^* \leftarrow$ BestOne$(S,S^*)$\\
		\STATE $S \leftarrow $ DiversifiedPerturb$(S, history)$ \\ 
	\ENDWHILE
 \end{algorithmic}
 \end{small}
 \end{algorithm}


\section{An iterated local search for Min-Diff DP}
\label{Sec:Hybrid Iterated Local Search}

\subsection{General framework}
\label{SubSec:General Framework}

Given a set of elements $N = \{e_1 ,e_2 ,\ldots,e_n\}$,  any subset set $S \subset N$ of $m$ elements is a legal or feasible  solution of \textsl{Min-Diff DP} and can be represented by $S = \{e_{S(1)},e_{S(2)},\ldots,e_{S(m)}\}$ ($1\leq S(i) \neq S(j) \leq n$ for all $i \neq j$) where $S(l) \ (1 \leq l \leq m)$ is the index of each selected element in $S$. 

Following the three-phase search framework \cite{Fu2015}, the proposed ILS\_MinDiff algorithm is composed of three main components: a descent-based neighborhood search procedure, a local optima exploring procedure and a local optima escaping procedure. Starting from a (good) initial solution provided by an initial solution generation procedure (Section \ref{SubSec:Initialization}), ILS\_MinDiff first employs the descent neighborhood search procedure to quickly attain a local optimal solution (Section \ref{SubSec:Descent-based Neighborhood Search}). Then it switches to the local optima exploring procedure which attempts to discover better local optima around the attained local optimum (Section \ref{SubSec:Local Optima Exploring}). Once no improved solution can be found (the search is located in a deep local optimum), ILS\_MinDiff tries to escape from the current search region  and jump to a new region with the help of a strong perturbation operation (Section \ref{SubSec:Local Optima Escaping}). During the search, the best solution encountered is recorded in $S^b$ and updated whenever it is needed. ILS\_MinDiff repeats the above three phases until a stopping condition (in our case, a time limit $t_{max}$) is reached (Alg. \ref{Algorithm:ILS MinDiff algorithm}). The composing procedures of ILS\_MinDiff are presented in the next subsections.

\begin{algorithm}
\begin{small}
 \caption{Framework of the ILS\_MinDiff algorithm for \textsl{Min-Diff DP}}
 \label{Algorithm:ILS MinDiff algorithm}
 \begin{algorithmic}[1]
 	\STATE \sf \textbf{Input}: a problem instance and the time limit  $t_{max}$
 	\STATE \textbf{Output}: the best solution $S^b$ found\\
 	\STATE $S \leftarrow$  generate\_initial\_solution$()$ \ /$*$ generate a good initial solution, Section \ref{SubSec:Initialization} $*$/\\
	\STATE $S^b \leftarrow S$ /$*$ record the best solution found so far in $S^b$ $*$/\\
	\WHILE{a stopping condition is not reached}
		\STATE $S \leftarrow $ descent\_based\_neighborhood\_search$(S)$ \ \ \ \ \ \ /$*$ local search, Section \ref{SubSec:Descent-based Neighborhood Search} $*$/\\
		\STATE $S \leftarrow $ local\_optima\_exploring$(S)$ \ \ /$*$ explore nearby local optima, Section \ref{SubSec:Local Optima Exploring} $*$/\\ 
// update the best solution found so far
		\IF{$f(S) < f(S^b)$} 
			\STATE $S^b \leftarrow S$ 
		\ENDIF
		\STATE $S \leftarrow $ local\_optima\_escaping$(S)$ \ \ \ \ \ /$*$ escape from local optima, Section \ref{SubSec:Local Optima Escaping} $*$/\\ 
	\ENDWHILE
	\STATE \textbf{output}($S^b$)
 \end{algorithmic}
 \end{small}
 \end{algorithm}

\subsection{Initialization}
\label{SubSec:Initialization}

The ILS\_MinDiff algorithm requires an initial solution to start its search. In general, the initial solution can be generated by any means (e.g., a random procedure or a greedy heuristic). In this work, the search starts from an elite solution of good quality, which is obtained in the following way. From a random solution $S \in \Omega$ (i.e., any subset of $m$ elements), we apply the descent-based neighborhood search procedure (Section \ref{SubSec:Descent-based Neighborhood Search}) to improve $S$ until a local optimum is reached. We repeat the process ten times to obtain ten local optimal solutions among which we select the best one (i.e., having the smallest objective value) as the initial solution. This procedure allows us to obtain an initial solution of relatively high quality. However, for instances with $n \geqslant 3000$, this initialization process becomes too time-consuming. Thus the algorithm simply uses a random solution, instead of an elite solution, to start its search.

\subsection{Descent-based neighborhood search phase}
\label{SubSec:Descent-based Neighborhood Search}

To obtain a local optimum from a given starting solution, a neighborhood search procedure is needed. In our case, we employ a simple and fast \textsl{descent\_based\_neighborhood\_search()} procedure. This search procedure iteratively makes transitions from the incumbent solution to a new neighbor solution according to a given neighborhood relation such that each transition leads necessarily to a better solution. This improvement process runs until no improving neighbor solution is available in the neighborhood, in which case the incumbent solution corresponds to a local optimum with respect to the neighborhood.

Two important issues to consider when designing such a search procedure are the definition of the neighborhood  and a technique for a fast evaluation of neighbor solutions. The neighborhood $Neighbor$ explored by the \textsl{descent\_based\_neighborhood\_search()} procedure is based on the $swap$ operation, which was used in previous studies \cite{Duarte2015,Mladenovic2016}. Given a solution $S$, we define $swap(p,q)$ as the move that exchanges an element $e_p \in S$ with an element $e_q \in N \setminus S$. Each $swap(p,q)$ brings about a variation $\Delta f(S,p,q)$ in the objective function $f$. Let $S'$ be the neighbor solution obtained by applying $swap(p,q)$ to the solution $S$, then the objective variation $\Delta f(S,p,q)$ (also called the move gain) is given by $\Delta f(S,p,q)=f(S')-f(S)$. Obviously, the size of this swap-based neighborhood is bound by $\mathcal{O}(m(n-m))$. 

To evaluate the neighborhood as fast as possible, we adopt a popular incremental evaluation technique \cite{Duarte2015,Lai2016,Mladenovic2016} to streamline the calculation of $\Delta f(S,p,q)$. Once a $swap(p,q)$ move is performed, only the elements related to $e_p$ and $e_q$ are needed to be considered. 
Before calculating $\Delta f(S,p,q)$ caused by a $swap(p,q)$ move, we first estimate the $\Delta$ value of each element $e_w$ in $S$ as follows: 

\begin{equation} \label{Equ:Gain Update}
\Delta'(e_w) = 
\begin{cases}
\Delta(e_w) - d_{wp} + d_{wq} & \forall e_w \in S \setminus \{e_p\} \\
\sum_{e_z \in S \setminus \{e_p\}}d_{qz} & e_w = e_q \\
\end{cases}
\end{equation} 

Therefore, with the $swap(p,q)$ operation, the objective value of the resulting neighbor solution $S' = S \setminus \{e_p\} \cup \{e_q\}$ can be conveniently calculated according to the following formula:

\begin{equation}
f(S') = \max_{e_w \in S'} \{ \Delta'(e_w)\}- \min_{e_w \in S'} \{ \Delta'(e_w)\}
\end{equation}

Correspondingly, the move gain of $swap(p,q)$ can be finally computed as:

\begin{equation}
\Delta f(S,p,q) = f(S') - f(S)
\end{equation}

With the help of this updating strategy, we can calculate $\Delta f(S,p,q)$ in $\mathcal{O}(m)$ because one only needs to check the $m-1$ elements adjacent to the removed element $e_p$ in $S$ and the $m-1$  elements adjacent to the added element $e_q$ in $S$.

To explore the neighborhood, the \textsl{descent\_based\_neighborhood\_search()} procedure uses the \textsl{best improvement strategy}. In other words, the best improving neighbor solution (with the smallest negative move gain) is selected at each iteration (ties are broken at random). After each solution transition, the search is resumed from the new incumbent solution. When no improving neighbor solution exists in the neighborhood, the incumbent solution is a local optimum. In this case, the \textsl{descent\_based\_neighborhood\_search()} procedure terminates and returns the last solution as its output. Finally, after each swap operation, the $\Delta$ value of each element in $S$ is updated, which is achieved in $\mathcal{O}(n)$.

\subsection{Local optima exploring phase}
\label{SubSec:Local Optima Exploring}

Obviously, the descent-based neighborhood search phase will quickly fall into a local optimum because it only accepts improving solutions during the search. To intensify the search around the attained local optimum and discover other nearby local optima of higher quality, we introduce the \textsl{local\_optima\_exploring()} procedure, which iterates through a moderate perturbation operation and the descent-based neighborhood search procedure (Algorithm \ref{Algorithm:Local Optima Explore}).

\begin{algorithm}
\begin{small}
 \caption{local\_optima\_exploring() procedure}
 \label{Algorithm:Local Optima Explore}
 \begin{algorithmic}[1]
 	\STATE \sf \textbf{Input}: a starting solution $S$ and the given search depth $nbr_{max}$
 	\STATE \textbf{Output}: the best solution $S^{*}$ found during the current local optima exploring phase\\
 	\STATE $S^{*} \leftarrow S$ \ \ \ \ \ \ \ \ \ \ \ \ \ \ \ \ \ \ \ \ \ \ \ \ \ \ \ \ \ \ \ \ \ \ \ \ \ \ \ \ /$*$ record the best solution found so far $*$/\\
	\STATE $nbr \leftarrow 0$
	\WHILE{$nbr < nbr_{max}$}
		\STATE $S \leftarrow $ weak\_perturb\_operator$(S)$ \ \ \ \ \ \ \ \ \ \ /$*$ perform a weak perturb operation $*$/\\
		\STATE $S \leftarrow $ descent\_based\_neighborhood\_search$(S)$ \ \ \ \ \ \ \ /$*$ attain a local optimum $*$/\\
		 // update the best solution found
		\IF{$f(S) < f(S^{*})$} 
			\STATE $S^{*} \leftarrow S$\\
			\STATE $nbr \leftarrow 0$\\
		\ELSE
			\STATE $nbr \leftarrow nbr + 1$\\
		\ENDIF
	\ENDWHILE
	\STATE \textbf{output}($S^{*}$)
\end{algorithmic}
\end{small}
\end{algorithm}

The \textsl{local\_optima\_exploring()} procedure starts by modifying slightly the input local optimal solution $S$ with the \textsl{weak\_perturb\_operator()}. At each perturbation step, we first generate at random $n+1$ neighbor solutions of the incumbent solution and then use the best one among these solutions to replace the incumbent solution. The \textsl{weak\_perturb\_operator()} repeats $p_w$ times ($p_w$ is a parameter called \textit{weak perturbation strength}), and returns the last perturbed solution which serves as the starting point of the descent-based neighborhood search. It is clear that a small (large) $p_w$ leads to a perturbed solution which is close to (far away from) the input solution $S$. In this work, we set $p_w = 2,3$.

Starting from the perturbed solution delivered by the \textsl{weak\_perturb\_operator()}, the \textsl{descent\_based\_neighborhood\_search()} procedure is run to attain a new local optimum, which becomes the incumbent solution of next iteration of the current local optima exploring phase. The best local optimum $S^*$ found during the local optima exploring phase is updated each time a new local optimum better than the recorded $S^*$ is encountered. The \textsl{local\_optima\_exploring()} procedure terminates when the recorded best local optimum $S^*$ cannot be updated for $nbr_{max}$ consecutive iterations ($nbr_{max}$ is a parameter called \textit{search depth}), indicating that the  region around the initial input solution $S$ is exhausted and the search needs to move into a more distant region, which is the purpose of the local optima escaping procedure described in the next section.

\subsection{Local optima escaping phase}
\label{SubSec:Local Optima Escaping}


To move away from the best local optimum $S^*$ found by \textsl{local\_optima\_exploring()}, we call for the \textsl{local\_optima\_escaping()} procedure which applies a strong perturbation mechanism. Specifically, the \textsl{local\_optima\_escaping()} procedure takes $S^*$ as its input, and then randomly performs $p_s$ swap operations. $p_s$, called strong perturbation strength, is defined by $p_s = \alpha \times n/m$, where $\alpha \in [1.0, 2.0)$ is a parameter called \textit{strong perturbation coefficient}. Since the objective variations are not considered during the perturbation operations, the perturbed solution may be greatly different from the input local optimum. This is particularly true with large $p_s$ value (e.g., $p_s > 10$), which definitively helps the search to jump into a distant search region.


\section{Computational experiments}
\label{Computational Experiments}

This section is dedicated to a performance assessment of the ILS\_MinDiff algorithm. For this purpose, we carry out an extensive experimental comparison between the proposed algorithm and the best-performing and the most recent VNS\_MinDiff algorithm \cite{Mladenovic2016} on six data sets of 190 benchmark instances. 

\subsection{Benchmark instances}
\label{SubSec:Benchmark Instances}

\textsl{MDPLIB}\footnote{\url{http://www.optsicom.es/mdp/mdplib_2010.zip}} proposes a comprehensive set of instances which are widely used for testing algorithms for solving diversity and dispersion problems. By excluding the small and easy instances, the remaining 190 benchmark instances tested in this work include the following three types and are classified into six data sets:
\begin{itemize}
\item \textbf{SOM} (\textsl{SOM-b}): This data set includes 20 test instances whose sizes range from $(n,m) = (100,10)$ to $(n,m) = (500,200)$. The instances of this set were created with a generator developed by Silva et al. \cite{Silva2004}.
\item \textbf{GKD} (\textsl{GKD-b} and \textsl{GKD-c}): These two data sets include 70 test instances whose sizes range from $(n,m) = (25,2)$ to $(n,m) = (500,50)$. The distance matrices are built by calculating the Euclidean distance between each pair of randomly generated points from the square $[0,10] \times [0,10]$. These instances were introduced by Glover et al. \cite{Glover1998} and generated by Duarte and Mart{\'\i} \cite{Duarte2007}, and Mart{\'\i} et al. \cite{Marti2010}.
\item \textbf{MDG} (\textsl{MDG-a}, \textsl{MDG-b} and \textsl{MDG-c}): The whole data set is composed of 100 test instances whose sizes range from $(n,m) = (500,50)$ to $(n,m) = (3000,600)$. The distance matrices in these instances are generated by selecting real numbers between 0 and 10 from an uniform distribution. These instances have been widely used in, e.g., Duarte and Mart{\'\i} \cite{Duarte2007}, Palubeckis \cite{Palubeckis2007}, and Mart{\'\i} et al. \cite{Marti2013}.
\end{itemize}

\subsection{Experimental settings}
\label{Experimental Settings}


The ILS\_MinDiff algorithm was implemented in C++ and compiled using g++ compiler with the `-O2' flag. All experiments were carried out on an Intel Xeon E5440 processor with 2.83 GHz and 2 GB RAM under Linux operating system. Without using any compiler flag, running the well-known DIMACS machine benchmark procedure dfmax.c\footnote{dfmax: \url{ftp://dimacs.rutgers.edu/pub/dsj/clique}} on our machine requires respectively 0.44, 2.63 and 9.85 seconds to solve the benchmark graphs r300.5, r400.5 and r500.5.

\begin{table*}[!htbp]
\caption{Parameter settings of the proposed ILS\_MinDiff algorithm}
\label{Tab:Parameters Table}
\vskip 0.15in
\begin{center}
\begin{scriptsize}
\begin{tabular}{lcll}
\toprule[0.75pt]
Parameters & Section & Description & Value\\
\midrule[0.5pt]
$t_{max}$   & \ref{SubSec:General Framework}      & time limit & $n$ \\
$nbr_{max}$ & \ref{SubSec:Local Optima Exploring} & search depth & 5\\
$p_w$       & \ref{SubSec:Local Optima Exploring} & weak perturbation strength & $\{2,3\}$\\
$\alpha$    & \ref{SubSec:Local Optima Escaping}  & strong perturbation coefficient & 1.0\\
\bottomrule[0.75pt]
\end{tabular}
\end{scriptsize}
\end{center}
\vskip -0.1in
\end{table*}

Given its stochastic nature, ILS\_MinDiff was independently executed, like \cite{Duarte2015,Mladenovic2016}, forty times with different random seeds on each test instance. Each run stops if the running time reaches the cut-off time limit ($t_{max}$). Following the literature \cite{Duarte2015,Mladenovic2016}, we set the time limit $t_{max}$ to $n$, where $n$ is the number of elements in the considered test instance. To run the ILS\_MinDiff algorithm, there are three parameters to be determined, including search depth $nbr_{max}$, weak perturbation strength $p_w$ in the local optima exploring phase and strong perturbation coefficient $\alpha$ in the local optima escaping phase. These parameters were fixed according to the experimental analysis of Section \ref{Sec:Experimental Analysis}: $nbr_{max}$ = 5 for instances of all six data sets; $p_w=3$ for instances with $n < 500$ or $n = 500, n/m < 10$, and $p_w=2$ for the remaining instances; $\alpha=1.0$ for all instances. A detailed description of the parameter settings is provided in Table \ref{Tab:Parameters Table}. It would be possible that fine tuning these parameters would lead to better results. As we show below, with the adopted parameter settings, ILS\_MinDiff already performs very well relative to the state-of-the-art results.  

\subsection{Comparison with the state-of-the-art algorithms}
\label{SubSec:Comparison With Other Algorithms}

As indicated in the introduction, three main approaches have been recently proposed in the literature to solve \textsl{Min-Diff DP}, including mixed integer programming (MIP) in 2009 \cite{Prokopyev2009}, greedy randomized adaptive search procedure with exterior path relinking (GRASP\_EPR)  in 2015 \cite{Duarte2015} and variable neighborhood search (VNS\_MinDiff) in 2016 \cite{Mladenovic2016}. It was shown in \cite{Mladenovic2016} that the latest VNS\_MinDiff algorithm performs the best by updating the best-known solutions for 170 out of 190 benchmark instances which were previously established by GRASP\_EPR of \cite{Duarte2015} while the exact MIP approach can only be applied to solve instances of small sizes (up to $n=40$ and $m=15$). Consequently, we adopt VNS\_MinDiff as the reference algorithm to assess the performance of the proposed ILS\_MinDiff algorithm. Detailed computational results of VNS\_MinDiff were extracted from \url{http://www.mi.sanu.ac.rs/~nenad/mddp/}. VNS\_MinDiff was coded in C++ and run on a computer with an Intel Core i7 2600 3.4 GHz CPU and 16 GB of RAM. Each instance was solved forty times with the time limit $t_{max} = n$ per run. As explained below, the computer used to run VNS\_MinDiff is roughly 1.2 times faster than our computer. So the stopping condition $t_{max} = n$ used by the two compared algorithms is more favorable for the reference VNS\_MinDiff algorithm than for our ILS\_MinDiff algorithm.

Given that the compared algorithms was executed on different platforms with different configurations, it seems difficult to strictly compare the runtimes. Therefore, we use \textit{solution quality} as the main criterion for our comparative studies. Nevertheless, we also report the CPU times consumed by the compared algorithms, which can still provide some useful indications about the computational efficiency of each algorithm. To make a meaningful comparison of the runtimes, we convert the CPU times reported for the reference algorithm with a scaling factor of 1.2 based on the frequencies of the two processors $(3.4/2.83 \approx 1.2)$, like previous studies \cite{Chen2016,Mei2014}. This linear conversion is based on the assumption that the CPU speed is approximately linearly proportional to the CPU frequency. Since the computing time of each algorithm is not only influenced by the frequency, but also by some other factors \cite{Henning2000}, the timing information was provided only for indicative purposes. 

The comparative results of the proposed ILS\_MinDiff algorithm and the reference VNS\_MinDiff algorithm are presented in Tables \ref{Tab:Comparisons on SOM-b}-\ref{Tab:Comparisons on MDG-c}. In these tables, column 1 gives the name of each instance (Instance), columns 2-5 and columns 6-9 respectively report the best objective value ($f_{best}$) obtained during forty runs, the average objective value ($f_{avg}$), the worst objective value ($f_{worst}$) and the corresponding average CPU time consumed ($t_{avg}$). For ILS\_MinDiff, we also report the standard deviation ($\sigma$), while this information is not available for the VNS\_MinDiff algorithm. The last column indicates the difference $\Delta f_{best}$ between the best solution values found by ILS\_MinDiff and VNS\_MinDiff (a negative value indicates an improved result). The best values among the results of these two algorithms are highlighted in bold. Note that the average CPU times of VNS\_MinDiff are scaled with the multiplication factor of 1.2 as explained previously. At the last two rows of each table, we also indicate the average value of each comparison indicator as well as the number of instances for which an algorithm shows a better performance compared to the other algorithm. An inapplicable entry was marked by ``$-$''.

To analyze the experimental results, we resort to the well-known \textsl{two-tailed sign test} \cite{Demsar2006} to check the significant difference on each comparison indicator between the compared algorithms. The two-tailed sign test is a popular technique to compare the overall performances of algorithms by counting the number of instances for which an algorithm is the overall winner. When two algorithms are compared, the corresponding null-hypothesis is that two algorithms are equivalent. The null-hypothesis is accepted if and only if each algorithm wins on approximately $X/2$ out of $X$ instances. Since tied matches support the null-hypothesis \cite{Demsar2006}, we split them evenly between the two compared algorithms, i.e., each one wins 0.5. The two-tailed sign test rejected the null hypothesis in all six data sets, suggesting these two algorithms do not have an equal performance. At a significance level of 0.05, the \textsl{Critical Values} ($CV$) of the two-tailed sign test were respectively $CV^{20}_{0.05} = 15$, $CV^{40}_{0.05} = 27$, and $CV^{50}_{0.05} = 32$ when the number of instances in each data set is $X = 20$, $X = 40$, and $X = 50$. This means that algorithm A is significantly better than algorithm B if A wins at least $CV^{X}_{0.05}$ instances for a data set of $X$ instances. 


\begin{table*}[!htbp]
\caption{Comparison of the results obtained by the reference VNS\_MinDiff algorithm and the proposed ILS\_MinDiff algorithm on the 20 instances of data set \textsl{SOM-b}.}
\label{Tab:Comparisons on SOM-b}
\vskip 0.15in
\begin{center}
\begin{tiny}
\begin{tabular}{l|rrrrcrrrrrr}
\toprule[0.75pt]
\multicolumn{1}{c}{} & \multicolumn{4}{c}{VNS\_MinDiff} && \multicolumn{5}{c}{ILS\_MinDiff} & \multicolumn{1}{c}{}\\
\cmidrule[0.5pt]{2-5} \cmidrule[0.5pt]{7-11}
Instance 	 & $f_{best}$ & $f_{avg}$ & $f_{worst}$ & $t_{avg}$ && $f_{best}$ & $f_{avg}$ & $f_{worst}$  & $t_{avg}$ & $\sigma$ & $\delta f_{best}$\\
\midrule[0.5pt]
SOM-b\_1\_n100\_m10   &1.00 &1.00 &\textbf{1.00} &\textbf{1.49}  &&\textbf{0.00} &\textbf{0.83} &\textbf{1.00} &12.06 &0.38&-1.00\\
SOM-b\_2\_n100\_m20   &\textbf{4.00} &4.50 &\textbf{5.00} &\textbf{28.56} &&\textbf{4.00} &\textbf{4.40} &\textbf{5.00} &29.62 &0.49& 0.00\\
SOM-b\_3\_n100\_m30   &8.00 &8.60 &\textbf{9.00} &\textbf{21.67} &&\textbf{7.00} &\textbf{7.88} &\textbf{9.00} &23.82 &0.40&-1.00\\
SOM-b\_4\_n100\_m40   &12.00&12.20&13.00&36.66 &&\textbf{10.00}&\textbf{10.98}&\textbf{12.00}&\textbf{34.41} &0.57&-2.00\\
SOM-b\_5\_n200\_m20   &\textbf{3.00} &\textbf{3.90} &\textbf{4.00} &82.22 &&\textbf{3.00} &4.00 &5.00 &\textbf{64.52} &0.39& 0.00\\
SOM-b\_6\_n200\_m40   &10.00&10.50&\textbf{11.00}&105.16 &&\textbf{9.00} &\textbf{10.40}&\textbf{11.00}&\textbf{47.47} &0.54&-1.00\\
SOM-b\_7\_n200\_m60   &16.00&16.70&18.00&90.11 &&\textbf{15.00}&\textbf{15.88}&\textbf{17.00}&\textbf{67.74} &0.51&-1.00\\
SOM-b\_8\_n200\_m80   &22.00&24.00&26.00&70.11 &&\textbf{19.00}&\textbf{21.50}&\textbf{23.00}&\textbf{76.80} &0.78&-3.00\\
SOM-b\_9\_n300\_m30   &7.00 &7.40 &\textbf{8.00} &99.61 &&\textbf{6.00} &\textbf{7.03} &\textbf{8.00} &\textbf{84.45} &0.42&-1.00\\
SOM-b\_10\_n300\_m60  &15.00&16.20&17.00&117.18 &&\textbf{14.00}&\textbf{15.55}&\textbf{16.00}&\textbf{103.76}&0.59&-1.00\\
SOM-b\_11\_n300\_m90  &22.00&24.10&26.00&112.84 &&\textbf{21.00}&\textbf{22.98}&\textbf{24.00}&\textbf{107.44}&0.72&-1.00\\
SOM-b\_12\_n300\_m120 &29.00&31.90&34.00&162.50 &&\textbf{28.00}&\textbf{30.13}&\textbf{32.00}&\textbf{110.24}&0.95&-1.00\\
SOM-b\_13\_n400\_m40  &10.00&10.40&11.00&\textbf{96.38} &&\textbf{9.00} &\textbf{9.85} &\textbf{10.00}&118.24&0.36&-1.00\\
SOM-b\_14\_n400\_m80  &\textbf{19.00}&21.30&23.00&198.88 &&\textbf{19.00}&\textbf{20.53}&\textbf{21.00}&\textbf{123.43}&0.55& 0.00\\
SOM-b\_15\_n400\_m120 &30.00&31.70&34.00&245.51 &&\textbf{28.00}&\textbf{30.03}&\textbf{32.00}&\textbf{160.33}&0.76&-2.00\\
SOM-b\_16\_n400\_m160 &40.00&43.40&47.00&306.05&&\textbf{36.00}&\textbf{38.73}&\textbf{41.00}&\textbf{188.42}&1.27&-4.00\\
SOM-b\_17\_n500\_m50  &12.00&12.80&\textbf{13.00}&235.04&&\textbf{11.00}&\textbf{12.33}&\textbf{13.00}&\textbf{158.23}&0.52&-1.00\\
SOM-b\_18\_n500\_m100 &\textbf{23.00}&25.10&27.00&278.50&&\textbf{23.00}&\textbf{24.70}&\textbf{26.00}&\textbf{238.64}&0.87& 0.00\\
SOM-b\_19\_n500\_m150 &36.00&39.60&45.00&297.66&&\textbf{34.00}&\textbf{37.00}&\textbf{38.00}&\textbf{224.23}&1.00&-2.00\\
SOM-b\_20\_n500\_m200 &49.00&56.40&63.00&322.28&&\textbf{43.00}&\textbf{47.10}&\textbf{50.00}&\textbf{264.56}&1.50&-6.00\\
\midrule[0.5pt]
Average value         &18.40&20.09&21.75&145.77&&\textbf{16.95}&\textbf{18.59}&\textbf{19.70}&\textbf{111.92}&0.68&-1.45\\
Wins	                  &2   &1     & 4   &4   &&18   &19   & 16  & 16 & -  & -   \\
\bottomrule[0.75pt]
\end{tabular}
\end{tiny}
\end{center}
\vskip -0.1in
\end{table*}

From Table \ref{Tab:Comparisons on SOM-b} which concerns the 20 instances of data set \textsl{SOM-b}, it can be observed that the proposed ILS\_MinDiff algorithm competes very favorably with the reference VNS\_MinDiff algorithm in all listed indicators. For this data set, ILS\_MinDiff easily dominates VNS\_MinDiff by improving the best-known results for 16 out of 20 instances and matching the best-known upper bounds for the remaining 4 instances. In addition, ILS\_MinDiff outperforms the reference algorithm in terms of the average solution value, the worst solution value, as well as the average CPU time. A small standard deviation over forty runs demonstrates that ILS\_MinDiff has a stable performance. In this case, $CV^{20}_{0.05} = 15$, the two-tailed sign test confirms the statistical significance of the differences between these two algorithms in all comparison indicators. 

\begin{table*}[!htbp]
\caption{Comparison of the results obtained by the reference VNS\_MinDiff algorithm and the proposed ILS\_MinDiff algorithm on the 50 instances of data set \textsl{GKD-b}.}
\label{Tab:Comparisons on GKD-b}
\vskip 0.15in
\begin{center}
\begin{tiny}
\begin{tabular}{l|rrrrcrrrrrr}
\toprule[0.75pt]
\multicolumn{1}{c}{} & \multicolumn{4}{c}{VNS\_MinDiff} && \multicolumn{5}{c}{ILS\_MinDiff} & \multicolumn{1}{c}{}\\
\cmidrule[0.5pt]{2-5} \cmidrule[0.5pt]{7-11}
Instance 	 & $f_{best}$ & $f_{avg}$ & $f_{worst}$ & $t_{avg}$ && $f_{best}$ & $f_{avg}$ & $f_{worst}$  & $t_{avg}$ & $\sigma$ & $\Delta f_{best}$\\
\midrule[0.5pt]
GKD-b\_1\_n25\_m2    &\textbf{0.00}  &\textbf{0.00}  &\textbf{0.00}  &\textbf{0.00} &&\textbf{0.00}  &\textbf{0.00}  &\textbf{0.00}  &\textbf{0.00} &0.00& 0.00\\
GKD-b\_2\_n25\_m2    &\textbf{0.00}  &\textbf{0.00}  &\textbf{0.00}  &\textbf{0.00} &&\textbf{0.00}  &\textbf{0.00}  &\textbf{0.00}  &\textbf{0.00} &0.00& 0.00\\
GKD-b\_3\_n25\_m2    &\textbf{0.00}  &\textbf{0.00}  &\textbf{0.00}  &\textbf{0.00} &&\textbf{0.00}  &\textbf{0.00}  &\textbf{0.00}  &\textbf{0.00} &0.00& 0.00\\
GKD-b\_4\_n25\_m2    &\textbf{0.00}  &\textbf{0.00}  &\textbf{0.00}  &\textbf{0.00} &&\textbf{0.00}  &\textbf{0.00}  &\textbf{0.00}  &\textbf{0.00} &0.00& 0.00\\
GKD-b\_5\_n25\_m2    &\textbf{0.00}  &\textbf{0.00}  &\textbf{0.00}  &\textbf{0.00} &&\textbf{0.00}  &\textbf{0.00}  &\textbf{0.00}  &\textbf{0.00} &0.00& 0.00\\
GKD-b\_6\_n25\_m7    &\textbf{12.72} &\textbf{12.72} &\textbf{12.72} &\textbf{0.00} &&\textbf{12.72} &\textbf{12.72} &\textbf{12.72} &0.12 &0.00& 0.00\\
GKD-b\_7\_n25\_m7    &\textbf{14.10} &\textbf{14.10} &\textbf{14.10} &\textbf{0.00} &&\textbf{14.10} &\textbf{14.10} &\textbf{14.10} &5.26 &0.00& 0.00\\
GKD-b\_8\_n25\_m7    &\textbf{16.76} &\textbf{16.76} &\textbf{16.76} &\textbf{0.00} &&\textbf{16.76} &\textbf{16.76} &\textbf{16.76} &6.37 &0.00& 0.00\\
GKD-b\_9\_n25\_m7    &\textbf{17.07} &\textbf{17.07} &\textbf{17.07} &\textbf{0.00} &&\textbf{17.07} &\textbf{17.07} &\textbf{17.07} &1.07 &0.00& 0.00\\
GKD-b\_10\_n25\_m7   &\textbf{23.27} &\textbf{23.27} &\textbf{23.27} &\textbf{0.00} &&\textbf{23.27} &\textbf{23.27} &\textbf{23.27} &5.72 &0.00& 0.00\\
GKD-b\_11\_n50\_m5   &\textbf{1.93}  &\textbf{1.93}  &\textbf{1.93}  &\textbf{0.01} &&\textbf{1.93}  &\textbf{1.93}  &\textbf{1.93}  &12.52&0.00& 0.00\\
GKD-b\_12\_n50\_m5   &\textbf{2.05}  &\textbf{2.05}  &\textbf{2.05}  &\textbf{0.05} &&\textbf{2.05}  &\textbf{2.05}  &\textbf{2.05}  &3.90 &0.00& 0.00\\
GKD-b\_13\_n50\_m5   &\textbf{2.36}  &\textbf{2.36}  &\textbf{2.36}  &\textbf{0.02} &&\textbf{2.36}  &\textbf{2.36}  &\textbf{2.36}  &1.79 &0.00& 0.00\\
GKD-b\_14\_n50\_m5   &\textbf{1.66}  &\textbf{1.66}  &\textbf{1.66}  &\textbf{0.01} &&\textbf{1.66}  &\textbf{1.66}  &\textbf{1.66}  &10.80&0.00& 0.00\\
GKD-b\_15\_n50\_m5   &\textbf{2.85}  &\textbf{2.85}  &\textbf{2.85}  &\textbf{0.00} &&\textbf{2.85}  &\textbf{2.85}  &\textbf{2.85}  &5.25 &0.00& 0.00\\
GKD-b\_16\_n50\_m15  &\textbf{42.75} &\textbf{42.75} &\textbf{42.75} &\textbf{0.01} &&\textbf{42.75} &\textbf{42.75} &\textbf{42.75} &13.74&0.00& 0.00\\
GKD-b\_17\_n50\_m15  &\textbf{48.11} &\textbf{48.11} &\textbf{48.11} &\textbf{0.00} &&\textbf{48.11} &\textbf{48.11} &\textbf{48.11} &8.83 &0.00& 0.00\\
GKD-b\_18\_n50\_m15  &\textbf{43.20} &\textbf{43.20} &\textbf{43.20} &\textbf{0.02} &&\textbf{43.20} &\textbf{43.20} &\textbf{43.20} &9.53 &0.00& 0.00\\
GKD-b\_19\_n50\_m15  &\textbf{46.41} &\textbf{46.41} &\textbf{46.41} &\textbf{0.30} &&\textbf{46.41} &\textbf{46.41} &\textbf{46.41} &13.40&0.00& 0.00\\
GKD-b\_20\_n50\_m15  &\textbf{47.72} &\textbf{47.72} &\textbf{47.72} &\textbf{0.00} &&\textbf{47.72} &\textbf{47.72} &\textbf{47.72} &16.15&0.00& 0.00\\
GKD-b\_21\_n100\_m10 &\textbf{9.33}  &9.38  &9.50  &39.18&&\textbf{9.33}  &\textbf{9.34}  &\textbf{9.43}  &\textbf{30.15}&0.02& 0.00\\
GKD-b\_22\_n100\_m10 &\textbf{8.04}  &9.01  &10.22 &46.02&&\textbf{8.04}  &\textbf{8.46}  &\textbf{9.58}  &\textbf{45.43}&0.41& 0.00\\
GKD-b\_23\_n100\_m10 &\textbf{6.91}  &7.63  &9.10  &\textbf{35.59}&&\textbf{6.91}  &\textbf{7.16}  &\textbf{8.64}  &52.55&0.42& 0.00\\
GKD-b\_24\_n100\_m10 &7.17  &7.65  &\textbf{7.94}  &\textbf{51.98}&&\textbf{5.79}  &\textbf{7.24}  &8.94  &53.52&0.76&-1.38\\
GKD-b\_25\_n100\_m10 &\textbf{6.91}  &8.65  &10.43 &\textbf{56.30}&&6.92  &\textbf{8.43}  &\textbf{10.28} &58.28&0.91& +0.01\\
GKD-b\_26\_n100\_m30 &\textbf{159.19}&160.37&166.31&\textbf{11.29} &&\textbf{159.19}&\textbf{159.19}&\textbf{159.19}&43.59&0.00& 0.00\\
GKD-b\_27\_n100\_m30 &\textbf{124.17}&125.05&127.10&\textbf{30.70}&&\textbf{124.17}&\textbf{124.17}&\textbf{124.17}&35.40&0.00& 0.00\\
GKD-b\_28\_n100\_m30 &\textbf{106.38}&\textbf{106.38}&\textbf{106.38}&\textbf{19.61}&&\textbf{106.38}&\textbf{106.38}&\textbf{106.38}&31.37&0.00& 0.00\\
GKD-b\_29\_n100\_m30 &\textbf{135.85}&\textbf{135.85}&\textbf{135.85}&\textbf{39.92}&&\textbf{135.85}&\textbf{135.85}&\textbf{135.85}&39.90&0.00& 0.00\\
GKD-b\_30\_n100\_m30 &\textbf{127.27}&\textbf{127.27}&\textbf{127.27}&\textbf{12.04}&&\textbf{127.27}&\textbf{127.27}&\textbf{127.27}&38.36&0.00& 0.00\\
GKD-b\_31\_n125\_m12 &\textbf{11.05} &\textbf{11.05} &\textbf{11.05} &\textbf{1.02} &&\textbf{11.05} &\textbf{11.05} &\textbf{11.05} &38.21&0.00& 0.00\\
GKD-b\_32\_n125\_m12 &11.60 &12.62 &\textbf{13.78} &67.16&&\textbf{10.36} &\textbf{12.41} &14.80 &\textbf{65.07}&1.07&-1.24\\
GKD-b\_33\_n125\_m12 &\textbf{9.18}  &11.47 &12.95 &\textbf{57.77}&&\textbf{9.18}  &\textbf{10.16} &\textbf{12.58} &63.48&0.88& 0.00\\
GKD-b\_34\_n125\_m12 &11.81 &\textbf{12.78} &\textbf{13.49} &74.72&&\textbf{10.79} &12.81 &14.37 &\textbf{59.28}&0.79&-1.02\\
GKD-b\_35\_n125\_m12 &\textbf{7.53}  &9.39  &11.69 &56.86&&\textbf{7.53}  &\textbf{8.72}  &\textbf{10.41} &\textbf{52.72}&0.91& 0.00\\
GKD-b\_36\_n125\_m37 &\textbf{125.55}&148.48&181.35&\textbf{38.64}&&\textbf{125.55}&\textbf{129.58}&\textbf{133.17}&59.51&1.79& 0.00\\
GKD-b\_37\_n125\_m37 &\textbf{194.22}&205.60&232.00&\textbf{39.10}&&\textbf{194.22}&\textbf{194.30}&\textbf{195.80}&48.05&0.35& 0.00\\
GKD-b\_38\_n125\_m37 &\textbf{184.27}&190.19&213.88&\textbf{20.03}&&\textbf{184.27}&\textbf{184.27}&\textbf{184.27}&48.78&0.00& 0.00\\
GKD-b\_39\_n125\_m37 &159.48&168.84&181.13&74.60&&\textbf{155.39}&\textbf{158.74}&\textbf{160.83}&\textbf{60.80}&1.60&-4.09\\
GKD-b\_40\_n125\_m37 &174.34&186.65&205.13&67.55&&\textbf{172.80}&\textbf{178.58}&\textbf{181.78}&\textbf{60.16}&1.63&-1.54\\
GKD-b\_41\_n150\_m15 &17.40 &19.60 &\textbf{20.98} &77.42&&\textbf{15.85} &\textbf{19.56} &22.44 &\textbf{69.33}&1.54&-1.55\\
GKD-b\_42\_n150\_m15 &18.20 &19.42 &\textbf{20.94} &88.09&&\textbf{13.96} &\textbf{17.96} &21.18 &\textbf{76.20}&1.63&-4.24\\
GKD-b\_43\_n150\_m15 &15.57 &17.66 &19.93 &69.96&&\textbf{11.83} &\textbf{16.56} &\textbf{19.56} &\textbf{62.96}&1.76&-3.74\\
GKD-b\_44\_n150\_m15 &15.16 &\textbf{17.58} &\textbf{19.28} &105.25&&\textbf{11.74} &17.73 &21.33 &\textbf{71.02}&1.87&-3.42\\
GKD-b\_45\_n150\_m15 &15.23 &18.38 &21.77 &78.11&&\textbf{12.84} &\textbf{17.83} &\textbf{20.50} &\textbf{75.93}&1.87&-2.39\\
GKD-b\_46\_n150\_m45 &\textbf{207.81}&225.89&258.53&\textbf{50.09}&&\textbf{207.81}&\textbf{209.38}&\textbf{214.56}&72.08&2.79& 0.00\\
GKD-b\_47\_n150\_m45 &214.42&228.62&259.86&\textbf{67.98}&&\textbf{211.77}&\textbf{214.14}&\textbf{218.14}&74.34&1.61&-2.65\\
GKD-b\_48\_n150\_m45 &180.00&221.04&239.52&\textbf{58.73}&&\textbf{177.29}&\textbf{184.01}&\textbf{192.36}&69.80&3.71&-2.71\\
GKD-b\_49\_n150\_m45 &205.39&227.51&288.07&\textbf{44.80}&&\textbf{197.88}&\textbf{201.12}&\textbf{206.75}&80.44&3.07&-7.51\\
GKD-b\_50\_n150\_m45 &\textbf{220.76}&245.17&279.55&96.04&&\textbf{220.76}&\textbf{225.78}&\textbf{234.81}&\textbf{78.84}&4.54& 0.00\\
\midrule[0.5pt]
Average value        &60.26 &64.36 &70.76 &\textbf{31.54}&&\textbf{59.51} &\textbf{60.82} &\textbf{62.27} &36.60&0.72&-0.75\\
Win                  &19    &14    &18    &34.5          &&31  &36    &32    &15.5 & -   & -    \\
\bottomrule[0.75pt]
\end{tabular}
\end{tiny}
\end{center}
\vskip -0.1in
\end{table*}

Table \ref{Tab:Comparisons on GKD-b} displays the comparative results of VNS\_MinDiff and ILS\_MinDiff on the 50 instances of data set \textsl{GKD-b}. As we can see from the table, the proposed ILS\_MinDiff algorithm achieves new best solutions (improved upper bounds) for 13 out of 50 instances and matches the best-known solutions for the remaining 37 instances. Moreover, compared to VNS\_MinDiff, ILS\_MinDiff respectively obtains a better average solution value and worst solution value for 36 and 32 out of 50 instances. For the average CPU time, ILS\_MinDiff wins much more instances than VNS\_MinDiff. It is worth mentioning that ILS\_MinDiff usually achieves a small standard deviation $\sigma$, and even $\sigma = 0.0$ for 27 out of 50 instances, which shows that ILS\_MinDiff  performs stably on this data set. Furthermore, the two-tailed sign test confirms that ILS\_MinDiff is significantly better than VNS\_MinDiff in terms of the average value and worst value, while for the best solution value, the number of instances wined by ILS\_MinDiff is only slightly smaller than the critical value, i.e., $31.5 < CV^{50}_{0.05} = 32$.

\begin{table*}[!htbp]
\caption{Comparison of the results obtained by the reference VNS\_MinDiff algorithm and the proposed ILS\_MinDiff algorithm on the 20 instances of data set \textsl{GKD-c}.}
\label{Tab:Comparisons on GKD-c}
\vskip 0.15in
\begin{center}
\begin{tiny}
\begin{tabular}{l|rrrrcrrrrrr}
\toprule[0.75pt]
\multicolumn{1}{c}{} & \multicolumn{4}{c}{VNS\_MinDiff} && \multicolumn{5}{c}{ILS\_MinDiff} & \multicolumn{1}{c}{}\\
\cmidrule[0.5pt]{2-5} \cmidrule[0.5pt]{7-11}
Instance 	 & $f_{best}$ & $f_{avg}$ & $f_{worst}$ & $t_{avg}$ && $f_{best}$ & $f_{avg}$ & $f_{worst}$  & $t_{avg}$ & $\sigma$ & $\Delta f_{best}$\\
\midrule[0.5pt]
GKD-c\_1\_n500\_m50  &10.09&12.09&13.63&413.95&&\textbf{9.85} &\textbf{10.85}&\textbf{11.93}&\textbf{221.70}&0.54&-0.24\\
GKD-c\_2\_n500\_m50  &10.79&12.98&15.41&421.76&&\textbf{10.06}&\textbf{11.64}&\textbf{12.79}&\textbf{202.54}&0.71&-0.73\\
GKD-c\_3\_n500\_m50  &\textbf{8.84} &11.81&16.29&439.88&&9.87 &\textbf{11.71}&\textbf{13.26}&\textbf{262.49}&0.68&+1.03\\
GKD-c\_4\_n500\_m50  &\textbf{9.42} &12.82&15.99&523.54&&9.55 &\textbf{11.51}&\textbf{13.60}&\textbf{262.46}&0.70&+0.13\\
GKD-c\_5\_n500\_m50  &11.26&13.53&16.00&282.88&&\textbf{10.36}&\textbf{12.03}&\textbf{13.11}&\textbf{267.60}&0.68&-0.90\\
GKD-c\_6\_n500\_m50  &10.63&12.14&14.49&440.82&&\textbf{10.21}&\textbf{11.36}&\textbf{12.37}&\textbf{254.12}&0.56&-0.42\\
GKD-c\_7\_n500\_m50  &11.58&13.71&16.65&\textbf{190.21}&&\textbf{10.47}&\textbf{12.27}&\textbf{13.45}&253.01&0.66&-1.11\\
GKD-c\_8\_n500\_m50  &11.31&13.79&20.50&409.92&&\textbf{9.89} &\textbf{11.55}&\textbf{12.88}&\textbf{287.74}&0.70&-1.42\\
GKD-c\_9\_n500\_m50  &10.45&13.71&17.64&384.04&&\textbf{9.69} &\textbf{11.29}&\textbf{12.61}&\textbf{262.99}&0.68&-0.76\\
GKD-c\_10\_n500\_m50 &\textbf{9.21} &13.94&18.45&430.34&&10.99&\textbf{12.14}&\textbf{13.24}&\textbf{288.20}&0.56&+1.78\\
GKD-c\_11\_n500\_m50 &11.03&12.39&14.59&444.01&&\textbf{9.03} &\textbf{10.99}&\textbf{12.10}&\textbf{206.79}&0.76&-2.00\\
GKD-c\_12\_n500\_m50 &\textbf{9.48} &12.98&15.06&364.58&&9.82 &\textbf{11.42}&\textbf{12.60}&\textbf{203.12}&0.63&+0.34\\
GKD-c\_13\_n500\_m50 &10.04&12.51&15.38&421.54&&\textbf{9.94} &\textbf{12.10}&\textbf{13.86}&\textbf{261.22}&0.78&-0.10\\
GKD-c\_14\_n500\_m50 &11.28&12.88&15.60&329.66&&\textbf{9.24} &\textbf{11.54}&\textbf{12.48}&\textbf{300.43}&0.62&-2.04\\
GKD-c\_15\_n500\_m50 &10.85&13.78&17.59&334.24&&\textbf{9.53} &\textbf{12.11}&\textbf{13.51}&\textbf{219.81}&0.88&-1.32\\
GKD-c\_16\_n500\_m50 &\textbf{8.39} &12.30&15.11&418.94&&10.04&\textbf{11.62}&\textbf{12.51}&\textbf{268.00}&0.58&+1.65\\
GKD-c\_17\_n500\_m50 &10.14&11.52&13.29&420.14&&\textbf{9.90} &\textbf{11.16}&\textbf{12.31}&\textbf{261.73}&0.60&-0.24\\
GKD-c\_18\_n500\_m50 &\textbf{9.77} &13.47&17.06&347.21&&10.56&\textbf{12.04}&\textbf{13.03}&\textbf{217.42}&0.56&+0.79\\
GKD-c\_19\_n500\_m50 &11.11&13.56&16.36&331.93&&\textbf{10.25}&\textbf{11.73}&\textbf{12.67}&\textbf{219.49}&0.62&-0.86\\
GKD-c\_20\_n500\_m50 &10.44&12.82&15.12&425.68&&\textbf{10.10}&\textbf{11.52}&\textbf{12.74}&\textbf{273.15}&0.62&-0.34\\
\midrule[0.5pt]
Average value        &10.31&12.94&16.01&388.76&&\textbf{9.97} &\textbf{11.63}&\textbf{12.85}&\textbf{249.70}&0.65&-0.34\\
Wins                 &6    &0    &0    &1     &&14   &20   &20   &19    & -   & -    \\
\bottomrule[0.75pt]
\end{tabular}
\end{tiny}
\end{center}
\vskip -0.1in
\end{table*}

Table \ref{Tab:Comparisons on GKD-c} presents the comparative results of VNS\_MinDiff and ILS\_MinDiff on the 20 instances of data set \textsl{GKD-c}. From Table \ref{Tab:Comparisons on GKD-c}, we observe that ILS\_MinDiff achieves a new improved solution for 14 out of 20 instances. ILS\_MinDiff fully dominates the reference algorithm by obtaining a better average solution value and worst solution value for all 20 instances. In addition, ILS\_MinDiff needs much less CPU times to achieve these results for almost all 20 instances. Although no significant difference is observed on the best solution values between ILS\_MinDiff and VNS\_MinDiff ($14 < CV^{20}_{0.05} = 15$), ILS\_MinDiff significantly performs better than VNS\_MinDiff in terms of the average solution value, the worst solution value and the average CPU time.

\begin{table*}[!htbp]
\caption{Comparison of the results obtained by the reference VNS\_MinDiff algorithm and the proposed ILS\_MinDiff algorithm on the 40 instances of data set \textsl{MDG-a}.}
\label{Tab:Comparisons on MDG-a}
\vskip 0.15in
\begin{center}
\begin{tiny}
\begin{tabular}{l|rrrrcrrrrrr}
\toprule[0.75pt]
\multicolumn{1}{c}{} & \multicolumn{4}{c}{VNS\_MinDiff} && \multicolumn{5}{c}{ILS\_MinDiff} & \multicolumn{1}{c}{}\\
\cmidrule[0.5pt]{2-5} \cmidrule[0.5pt]{7-11}
Instance 	 & $f_{best}$ & $f_{avg}$ & $f_{worst}$ & $t_{avg}$ && $f_{best}$ & $f_{avg}$ & $f_{worst}$  & $t_{avg}$ & $\sigma$ & $\Delta f_{best}$\\
\midrule[0.5pt]
MDG-a\_1\_n500\_m50    &11.34&12.26&\textbf{12.67}&\textbf{182.21} &&\textbf{11.11}&\textbf{12.09}&12.99&259.36 &0.43&-0.23\\
MDG-a\_2\_n500\_m50    &11.67&12.45&12.94&\textbf{226.92} &&\textbf{11.00}&\textbf{12.10}&\textbf{12.86}&\textbf{227.06} &0.40&-0.67\\
MDG-a\_3\_n500\_m50    &11.71&12.22&12.82&373.98 &&\textbf{11.31}&\textbf{12.08}&\textbf{12.68}&\textbf{253.43} &0.32&-0.40\\
MDG-a\_4\_n500\_m50    &11.56&12.34&12.94&310.02 &&\textbf{11.35}&\textbf{12.05}&\textbf{12.65}&\textbf{253.37} &0.34&-0.21\\
MDG-a\_5\_n500\_m50    &12.05&12.50&12.77&280.31 &&\textbf{10.75}&\textbf{12.08}&\textbf{12.67}&\textbf{275.97} &0.41&-1.30\\
MDG-a\_6\_n500\_m50    &\textbf{10.87}&12.15&12.74&394.31 &&11.11&\textbf{12.07}&\textbf{12.72}&\textbf{255.47} &0.40&+0.24\\
MDG-a\_7\_n500\_m50    &10.95&12.14&13.17&351.23 &&\textbf{10.64}&\textbf{12.08}&\textbf{12.7}6&\textbf{234.95} &0.46&-0.31\\
MDG-a\_8\_n500\_m50    &11.80&12.41&\textbf{13.00}&\textbf{245.36} &&\textbf{10.16}&\textbf{11.99}&\textbf{13.00}&251.16 &0.52&-1.64\\
MDG-a\_9\_n500\_m50    &11.54&12.37&12.80&248.57 &&\textbf{10.98}&\textbf{11.98}&\textbf{12.61}&\textbf{250.43} &0.40&-0.56\\
MDG-a\_10\_n500\_m50   &11.60&12.33&13.00&\textbf{298.28} &&\textbf{11.04}&\textbf{12.01}&\textbf{12.57}&252.52 &0.36&-0.56\\
MDG-a\_11\_n500\_m50   &11.25&12.12&\textbf{12.68}&\textbf{141.30} &&\textbf{10.83}&\textbf{11.96}&12.71&237.47 &0.47&-0.42\\
MDG-a\_12\_n500\_m50   &12.17&12.53&12.87&302.29 &&\textbf{11.12}&\textbf{12.03}&\textbf{12.69}&\textbf{244.0}3 &0.37&-1.05\\
MDG-a\_13\_n500\_m50   &12.05&12.41&12.99&358.21 &&\textbf{10.63}&\textbf{12.00}&\textbf{12.74}&\textbf{216.85} &0.46&-1.42\\
MDG-a\_14\_n500\_m50   &11.60&12.42&13.06&\textbf{197.84} &&\textbf{10.77}&\textbf{11.94}&\textbf{12.79}&289.92 &0.41&-0.83\\
MDG-a\_15\_n500\_m50   &11.55&12.39&12.91&\textbf{265.38} &&\textbf{10.97}&\textbf{12.01}&\textbf{12.77}&274.32 &0.42&-0.58\\
MDG-a\_16\_n500\_m50   &12.15&12.64&13.12&288.91 &&\textbf{10.65}&\textbf{12.04}&\textbf{12.76}&\textbf{242.01} &0.43&-1.50\\
MDG-a\_17\_n500\_m50   &11.76&12.32&\textbf{12.73}&331.67 &&\textbf{10.88}&\textbf{12.12}&12.94&\textbf{293.07} &0.45&-0.88\\
MDG-a\_18\_n500\_m50   &11.95&12.42&12.90&381.47 &&\textbf{10.88}&\textbf{12.12}&\textbf{12.60}&\textbf{286.69} &0.35&-1.07\\
MDG-a\_19\_n500\_m50   &\textbf{11.50}&\textbf{12.34}&12.93&289.75 &&11.57&\textbf{12.21}&12.71&\textbf{249.01} &0.28&+0.07\\
MDG-a\_20\_n500\_m50   &11.66&12.18&\textbf{12.60}&304.18 &&\textbf{11.10}&\textbf{12.17}&13.20&\textbf{284.12} &0.42&-0.56\\
MDG-a\_21\_n2000\_m200 &\textbf{50.00}&\textbf{53.10}&\textbf{57.00}&1631.33&&\textbf{50.00}&53.43&\textbf{57.00}&\textbf{1342.38}&1.79& 0.00\\
MDG-a\_22\_n2000\_m200 &51.00&53.60&\textbf{56.00}&1788.64&&\textbf{50.00}&\textbf{53.55}&58.00&\textbf{1446.83}&2.20&-1.00\\
MDG-a\_23\_n2000\_m200 &52.00&54.30&\textbf{57.00}&\textbf{1151.98} &&\textbf{49.00}&\textbf{53.60}&58.00&1380.25&2.49&-3.00\\
MDG-a\_24\_n2000\_m200 &\textbf{48.00}&\textbf{53.00}&\textbf{58.00}&1617.97&&50.00&53.63&\textbf{58.00}&\textbf{1447.23}&2.03&+2.00\\
MDG-a\_25\_n2000\_m200 &51.00&54.30&\textbf{58.00}&1506.10&&\textbf{50.00}&\textbf{53.60}&\textbf{58.00}&\textbf{1420.26}&1.95&-1.00\\
MDG-a\_26\_n2000\_m200 &\textbf{49.00}&\textbf{53.00}&\textbf{57.00}&\textbf{1363.76}&&50.00&53.58&\textbf{57.00}&1410.15&1.69&+1.00\\
MDG-a\_27\_n2000\_m200 &50.00&54.70&\textbf{58.00}&1435.36&&\textbf{49.00}&\textbf{53.73}&\textbf{58.00}&\textbf{1406.19}&2.18&-1.00\\
MDG-a\_28\_n2000\_m200 &\textbf{48.00}&53.10&\textbf{57.00}&1536.53&&50.00&\textbf{52.98}&59.00&\textbf{1464.31}&2.03&+2.00\\
MDG-a\_29\_n2000\_m200 &51.00&\textbf{53.00}&\textbf{56.00}&\textbf{1317.26}&&\textbf{47.00}&53.48&58.00&1542.65&2.61&-4.00\\
MDG-a\_30\_n2000\_m200 &50.00&\textbf{54.00}&\textbf{57.00}&\textbf{965.94} &&\textbf{49.00}&54.28&59.00&1469.70&2.93&-1.00\\
MDG-a\_31\_n2000\_m200 &50.00&54.50&60.00&\textbf{1301.65}&&\textbf{49.00}&\textbf{53.88}&\textbf{58.00}&1517.10&2.32&-1.00\\
MDG-a\_32\_n2000\_m200 &51.00&54.60&61.00&\textbf{1258.96}&&\textbf{48.00}&\textbf{53.25}&\textbf{57.00}&1383.72&1.84&-3.00\\
MDG-a\_33\_n2000\_m200 &49.00&\textbf{53.70}&60.00&\textbf{1578.54}&&\textbf{48.00}&53.80&\textbf{59.00}&1634.04&2.66&-1.00\\
MDG-a\_34\_n2000\_m200 &\textbf{49.00}&\textbf{53.30}&\textbf{57.00}&\textbf{1270.10}&&\textbf{49.00}&53.48&59.00&1388.50&2.23& 0.00\\
MDG-a\_35\_n2000\_m200 &53.00&54.70&\textbf{56.00}&\textbf{1224.68}&&\textbf{48.00}&\textbf{54.08}&59.00&1431.79&2.18&-5.00\\
MDG-a\_36\_n2000\_m200 &51.00&54.10&\textbf{57.00}&1560.10&&\textbf{48.00}&\textbf{53.73}&59.00&\textbf{1439.09}&2.41&-3.00\\
MDG-a\_37\_n2000\_m200 &49.00&\textbf{52.90}&\textbf{56.00}&1553.54&&\textbf{48.00}&53.85&58.00&\textbf{1398.92}&2.37&-1.00\\
MDG-a\_38\_n2000\_m200 &\textbf{48.00}&\textbf{53.60}&\textbf{57.00}&1515.90&&49.00&53.83&58.00&\textbf{1505.69}&2.24&+1.00\\
MDG-a\_39\_n2000\_m200 &51.00&54.00&\textbf{58.00}&1480.96&&\textbf{48.00}&\textbf{53.48}&\textbf{58.00}&\textbf{1451.99}&2.36&-3.00\\
MDG-a\_40\_n2000\_m200 &50.00&\textbf{53.20}&\textbf{56.00}&\textbf{1268.04}&&\textbf{49.00}&54.03&59.00&1361.78&2.30&-1.00\\
\midrule[0.5pt]
Average value          &30.84&33.04&\textbf{35.17}&851.83 &&\textbf{29.92}&\textbf{32.86}&35.49&\textbf{849.34} &1.32&-0.92\\
Wins                   &7    &10   &19   &16.5   &&33   &30   &21   &23.5   & -   & -    \\
\bottomrule[0.75pt]
\end{tabular}
\end{tiny}
\end{center}
\vskip -0.1in
\end{table*}

Table \ref{Tab:Comparisons on MDG-a} describes the comparative results of VNS\_MinDiff and ILS\_MinDiff on the 40 instances of data set \textsl{MDG-a}. This table shows that ILS\_MinDiff again outperforms the reference algorithm. Specifically, in terms of the best solution value, ILS\_MinDiff finds a new best solution for 32 out of 40 instances and matches the best-known solutions for 2 instances. Concerning the average solution value, ILS\_MinDiff remains competitive, achieving a better average solution value for 30 out of 40 instances. In addition, ILS\_MinDiff also wins much more instances than VNS\_MinDiff, i.e., 21 and 23.5 instances for the worst solution value and average CPU time respectively. With the two-tailed sign test, at a significance level of 0.05, we find that ILS\_MinDiff is significantly better than VNS\_MinDiff algorithm in terms of the best solution value ($33 > CV^{40}_{0.05} = 27$) and average solution value ($30 > CV^{40}_{0.05} = 27$).

\begin{table*}[!htbp]
\caption{Comparison of the results obtained by the reference VNS\_MinDiff algorithm and the proposed ILS\_MinDiff algorithm on the 40 instances of data set \textsl{MDG-b}.}
\label{Tab:Comparisons on MDG-b}
\vskip 0.15in
\begin{center}
\begin{tiny}
\begin{tabular}{l|rrrrcrrrrrr}
\toprule[0.75pt]
\multicolumn{1}{c}{} & \multicolumn{4}{c}{VNS\_MinDiff} && \multicolumn{5}{c}{ILS\_MinDiff} & \multicolumn{1}{c}{}\\
\cmidrule[0.5pt]{2-5} \cmidrule[0.5pt]{7-11}
Instance 	 & $f_{best}$ & $f_{avg}$ & $f_{worst}$ & $t_{avg}$ && $f_{best}$ & $f_{avg}$ & $f_{worst}$  & $t_{avg}$ & $\sigma$ & $\Delta f_{best}$\\
\midrule[0.5pt]
MDG-b\_1\_n500\_m50    &1185.11&1246.78&1296.49&319.40 &&\textbf{1118.48}&\textbf{1209.70}&\textbf{1265.10}&\textbf{198.74} &35.72 &-66.63 \\
MDG-b\_2\_n500\_m50    &1182.48&1256.77&1322.03&\textbf{294.17} &&\textbf{1082.03}&\textbf{1199.86}&\textbf{1249.69}&279.30 &35.52 &-100.45\\
MDG-b\_3\_n500\_m50    &\textbf{1070.87}&1243.84&1310.09&397.45 &&1104.07&\textbf{1203.60}&\textbf{1265.66}&\textbf{237.08} &39.72 &+33.20 \\
MDG-b\_4\_n500\_m50    &1153.93&1240.57&1287.46&287.94 &&\textbf{1052.27}&\textbf{1193.79}&\textbf{1258.01}&\textbf{226.04} &43.67 &-101.66\\
MDG-b\_5\_n500\_m50    &1209.80&1262.90&1317.82&\textbf{223.27} &&\textbf{1051.91}&\textbf{1204.83}&\textbf{1275.55}&225.86 &48.06 &-157.89\\
MDG-b\_6\_n500\_m50    &1071.61&1227.71&1319.86&358.58 &&\textbf{1061.50}&\textbf{1202.49}&\textbf{1292.94}&\textbf{242.20} &41.70 &-10.11 \\
MDG-b\_7\_n500\_m50    &1099.68&1215.38&1311.55&307.58 &&\textbf{1076.57}&\textbf{1205.87}&\textbf{1260.31}&\textbf{229.67} &38.66 &-23.11 \\
MDG-b\_8\_n500\_m50    &1185.59&1245.45&1316.97&296.41 &&\textbf{1005.45}&\textbf{1207.23}&\textbf{1278.51}&\textbf{250.86} &47.25 &-180.14\\
MDG-b\_9\_n500\_m50    &1154.33&1232.61&1261.83&292.68 &&\textbf{1116.63}&\textbf{1210.19}&\textbf{1260.14}&\textbf{269.01} &35.30 &-37.70 \\
MDG-b\_10\_n500\_m50   &1198.08&1242.15&1289.55&326.46 &&\textbf{1092.25}&\textbf{1195.55}&\textbf{1264.05}&\textbf{237.71} &44.78 &-105.83\\
MDG-b\_11\_n500\_m50   &1145.73&1221.54&1275.68&299.57 &&\textbf{1090.59}&\textbf{1202.24}&\textbf{1256.12}&\textbf{236.51} &36.85 &-55.14 \\
MDG-b\_12\_n500\_m50   &1165.43&1238.15&1294.60&303.54 &&\textbf{1093.89}&\textbf{1206.29}&\textbf{1287.75}&\textbf{232.29} &39.64 &-71.54 \\
MDG-b\_13\_n500\_m50   &1180.43&1238.00&\textbf{188.52}&\textbf{157.10} &&\textbf{1123.26}&\textbf{1205.87}&1296.97&268.81 &37.48 &-57.17 \\
MDG-b\_14\_n500\_m50   &1166.81&1247.25&1315.79&\textbf{180.78} &&\textbf{1089.38}&\textbf{1200.78}&\textbf{1258.22}&264.85 &36.19 &-77.43 \\
MDG-b\_15\_n500\_m50   &1220.83&1273.74&1314.10&337.81 &&\textbf{1108.10}&\textbf{1207.73}&\textbf{1287.25}&\textbf{231.45} &41.54 &-112.73\\
MDG-b\_16\_n500\_m50   &1176.16&1248.31&1317.30&354.16 &&\textbf{1132.34}&\textbf{1217.24}&\textbf{1266.44}&\textbf{313.70} &31.32 &-43.82 \\
MDG-b\_17\_n500\_m50   &1174.66&1252.52&1297.05&383.77 &&\textbf{1062.69}&\textbf{1200.76}&\textbf{1275.49}&\textbf{228.43} &44.04 &-111.97\\
MDG-b\_18\_n500\_m50   &1187.82&1267.94&1338.04&\textbf{182.72} &&\textbf{1032.54}&\textbf{1198.20}&\textbf{1271.48}&253.42 &45.10 &-155.28\\
MDG-b\_19\_n500\_m50   &1175.26&1257.81&1291.88&443.22 &&\textbf{1076.27}&\textbf{1206.94}&\textbf{1272.92}&\textbf{269.22} &39.22 &-98.99 \\
MDG-b\_20\_n500\_m50   &1151.34&1233.04&1285.53&326.04 &&\textbf{1050.33}&\textbf{1199.99}&\textbf{1260.42}&\textbf{242.06} &42.23 &-101.01\\
MDG-b\_21\_n2000\_m200 &4083.16&4468.62&4737.59&\textbf{1230.96}&&\textbf{3978.52}&\textbf{4299.38}&\textbf{4552.68}&1366.04&138.11&-104.64\\
MDG-b\_22\_n2000\_m200 &4187.77&4540.42&4952.63&\textbf{1247.75}9&&\textbf{3911.34}&\textbf{4377.97}&\textbf{4678.26}&1466.30&178.38&-276.43\\
MDG-b\_23\_n2000\_m200 &4237.38&4489.07&5171.39&\textbf{1592.87}&&\textbf{4127.34}&\textbf{4422.12}&\textbf{4858.86}&1591.88&183.29&-110.04\\
MDG-b\_24\_n2000\_m200 &4212.28&4452.33&4708.87&\textbf{1202.50}&&\textbf{4088.26}&\textbf{4421.77}&\textbf{4705.92}&1513.38&147.62&-124.02\\
MDG-b\_25\_n2000\_m200 &4152.88&4435.36&4713.25&1650.97&&\textbf{3892.67}&\textbf{4340.78}&\textbf{4577.78}&\textbf{1463.02}&132.66&-260.21\\
MDG-b\_26\_n2000\_m200 &\textbf{4039.92}&4497.39&4798.83&1581.28&&4116.90&\textbf{4423.07}&\textbf{4775.00}&\textbf{1535.76}&147.65&+76.98 \\
MDG-b\_27\_n2000\_m200 &\textbf{4010.77}&4486.90&4855.86&\textbf{1295.65}&&4126.90&\textbf{4424.59}&\textbf{4711.53}&1599.67&150.02&+116.13\\
MDG-b\_28\_n2000\_m200 &4206.07&4498.25&\textbf{4798.32}&\textbf{1013.33} &&\textbf{4112.43}&\textbf{4446.16}&4975.79&1579.61&173.71&-93.64 \\
MDG-b\_29\_n2000\_m200 &4214.79&4505.51&\textbf{4809.00}&\textbf{1244.74}&&\textbf{4057.62}&\textbf{4377.08}&4846.05&1450.96&164.65&-157.17\\
MDG-b\_30\_n2000\_m200 &4272.07&4564.38&\textbf{4786.12}&\textbf{1227.43}&&\textbf{4110.61}&\textbf{4470.64}&4885.52&1519.61&178.31&-161.47\\
MDG-b\_31\_n2000\_m200 &4328.97&4474.43&4710.96&1498.39&&\textbf{4074.80}&\textbf{4323.11} &\textbf{4539.19}&\textbf{1155.72}&120.41&-254.17\\
MDG-b\_32\_n2000\_m200 &4226.55&4484.07&4664.58&1283.56&&\textbf{3929.49}&\textbf{4301.35}&\textbf{4628.75}&\textbf{1278.11}&159.57&-297.06\\
MDG-b\_33\_n2000\_m200 &4037.50&4387.64&4786.52&1538.08&&\textbf{3985.32}&\textbf{4351.01}&\textbf{4624.97}&\textbf{1396.08}&148.77&-52.18 \\
MDG-b\_34\_n2000\_m200 &4279.58&4480.58&\textbf{4850.85}&\textbf{1245.97}&&\textbf{4084.46}&\textbf{4402.11}&4965.12&1611.59&183.27&-195.12\\
MDG-b\_35\_n2000\_m200 &4018.60&\textbf{4367.05}&\textbf{4679.23}&1899.08&&\textbf{4000.31}&4396.43&4824.11&\textbf{1432.83}&163.53&-18.30 \\
MDG-b\_36\_n2000\_m200 &4231.38&\textbf{4433.05}&\textbf{4674.14}&\textbf{1281.22}&&\textbf{4095.13}&4435.33&4955.35&1678.21&172.50&-136.25\\
MDG-b\_37\_n2000\_m200 &4100.54&4472.45&\textbf{4834.64}&1774.86&&\textbf{4035.74}&\textbf{4409.06}&4873.27&\textbf{1524.78}&164.91&-64.80 \\
MDG-b\_38\_n2000\_m200 &4136.67&4506.89&\textbf{4802.26}&1515.20&&\textbf{4126.69}&\textbf{4418.53}&4817.94&\textbf{1443.40}&171.30&-9.98  \\
MDG-b\_39\_n2000\_m200 &4242.30&4450.95&\textbf{4635.06}&\textbf{1462.99}&&\textbf{4131.87}&\textbf{4403.46}&4807.67&1486.56&137.81&-110.43\\
MDG-b\_40\_n2000\_m200 &4249.76&4556.52&4804.78&1706.47&&\textbf{4306.02}&\textbf{4306.02}&\textbf{4549.61}&\textbf{1232.07}&138.38&-226.72\\
\midrule[0.5pt]
Average value          &2668.12&2861.11&3045.47&864.93 &&\textbf{2565.75}&\textbf{2795.73}&\textbf{3013.91}&\textbf{856.57} &98.97 &-102.37\\
Wins                   &3      &2      &10     &14.5     &&37     &38     &30     &25.5     & - & -      \\
\bottomrule[0.75pt]
\end{tabular}
\end{tiny}
\end{center}
\vskip -0.1in
\end{table*}

Table \ref{Tab:Comparisons on MDG-b} on the 40 instances of data set \textsl{MDG-b} demonstrates that ILS\_MinDiff achieves a highly competitive performance by obtaining a new best solution for 37 out of 40 instances. Moreover, ILS\_MinDiff achieves a better average solution value and worst solution value, and respectively wins 38 and 30 out of 40 instances with respect to VNS\_MinDiff. The two-tailed sign test confirms that the differences of these two algorithms are statistically significant. ILS\_MinDiff uses less time than VNS\_MinDiff for 26 out of 40 instances. It is only sightly smaller than the critical value of the statistical test at the significance level of 0.05, i.e., $26 < CV^{40}_{0.05} = 27$. It is worth noting that the ILS\_MinDiff algorithm achieves these results with a relatively large average standard deviation 98.97, suggesting that ILS\_MinDiff is less stable on this data set than on the previous data sets. 

Table \ref{Tab:Comparisons on MDG-c} on the 20 instances of data set \textsl{MDG-c} shows that the ILS\_MinDiff algorithm obtains a new best solution for all 20 instances except for \textsl{MDG-c\_18\_n3000\_m600}. In terms of the best solution value, ILS\_MinDiff is significantly better than the reference algorithm, i.e., $19 > CV^{20}_{0.05} = 15$. Compared with VNS\_MinDiff, we also observe that ILS\_MinDiff achieves a competitive performance in terms of the average solution value and the worst solution value, and respectively wins 11 and 9 out of 20 instances. VNS\_MinDiff has a better performance on the last 9 out of 10 instances in terms of the average solution value and the worst solution value, but consumes much more time than ILS\_MinDiff\footnote{We observe that VNS\_MinDiff reports abnormal computing times for several instances, exceeding the allowed time limit ($t_{max}=n$ scaled with 1.2, i.e., 3600 seconds). No information is available to explain these anomalies.}. Finally, to verify if ILS\_MinDiff can improve its results on these 9 instances, we ran ILS\_MinDiff with a relaxed time limit of $3000 \times 1.2 = 3600$ seconds, which corresponds to the time limit of VNS\_MinDiff. ILS\_MinDiff actually achieved better results than VNS\_MinDiff not only in terms of the best solution value, the average solution value as well the worst solution value. For example, the best, average, worst solutions of ILS\_MinDiff for \textsl{MDG-c\_11\_n3000\_m500} are respectively improved to 10295.00, 11123.25 and 12214.00, and the average CPU time is 3237.36 seconds. Similarly, we also achieved better results for \textsl{MDG-c\_12\_n3000\_m500}, i.e., $f_{best} = 9909.00$, $f_{avg} = 10894.98$, $f_{worst} = 12328.00$ and $t_{avg} = 3132.11$.

\begin{sidewaystable}
\caption{Comparison of the results obtained by the reference VNS\_MinDiff algorithm and the proposed ILS\_MinDiff algorithm on the 20 instances of data set \textsl{MDG-c}.}
\label{Tab:Comparisons on MDG-c}
\vskip 0.15in
\begin{center}
\begin{tiny}
\begin{tabular}{l|rrrrcrrrrrr}
\toprule[0.75pt]
\multicolumn{1}{c}{} & \multicolumn{4}{c}{VNS\_MinDiff} && \multicolumn{5}{c}{ILS\_MinDiff} & \multicolumn{1}{c}{}\\
\cmidrule[0.5pt]{2-5} \cmidrule[0.5pt]{7-11}
Instance 	 & $f_{best}$ & $f_{avg}$ & $f_{worst}$ & $t_{avg}$ && $f_{best}$ & $f_{avg}$ & $f_{worst}$  & $t_{avg}$ & $\sigma$ & $\Delta f_{best}$\\
\midrule[0.5pt]
MDG-c\_1\_n3000\_m300  &6344.00 &6595.40 &7135.00 &\textbf{1746.54}&&\textbf{5772.00} &\textbf{6265.60} &\textbf{6794.00} &2322.74&290.56 &-572.00\\
MDG-c\_2\_n3000\_m300  &6109.00 &6651.50 &7183.00 &\textbf{1584.34}&&\textbf{5936.00} &\textbf{6539.33} &\textbf{7114.00} &2772.71&300.02 &-173.00\\
MDG-c\_3\_n3000\_m300  &6365.00 &6828.70 &7221.00 &\textbf{1967.66}&&\textbf{5585.00} &\textbf{6243.03} &\textbf{7006.00} &2501.82&277.13 &-780.00\\
MDG-c\_4\_n3000\_m300  &6304.00 &6787.10 &7215.00 &\textbf{1553.74}&&\textbf{5969.00} &\textbf{6636.75} &\textbf{7168.00} &2841.41&299.52 &-335.00\\
MDG-c\_5\_n3000\_m300  &5954.00 &6729.30 &\textbf{7282.00} &\textbf{1977.83}&&\textbf{5750.00} &\textbf{6663.25} &7405.00 &2750.35&343.32 &-204.00\\
MDG-c\_6\_n3000\_m400  &8403.00 &9422.10 &10592.00&\textbf{2233.43}&&\textbf{7648.00} &\textbf{8412.98} &\textbf{8999.00} &2781.03&334.32 &-755.00\\
MDG-c\_7\_n3000\_m400  &8606.00 &9308.60 &9770.00 &\textbf{2216.87}&&\textbf{7829.00} &\textbf{8457.15} &\textbf{9522.00} &2837.35&373.83 &-777.00\\
MDG-c\_8\_n3000\_m400  &8217.00 &9206.80 &10219.00&\textbf{2411.77}&&\textbf{7984.00} &\textbf{8497.28} &\textbf{9033.00} &2699.22&240.93 &-233.00\\
MDG-c\_9\_n3000\_m400  &8478.00 &9140.50 &10337.00&\textbf{2499.54}&&\textbf{7657.00} &\textbf{8259.35} &\textbf{9128.00} &2626.88&371.85 &-821.00\\
MDG-c\_10\_n3000\_m400 &8244.00 &9372.30 &\textbf{10129.00}&\textbf{2186.04}&&\textbf{7672.00} &\textbf{8646.00} &10432.00&2992.54&589.01 &-572.00\\
MDG-c\_11\_n3000\_m500 &11145.00&\textbf{11998.90}&\textbf{13151.00}&3641.36&&\textbf{11031.00}&12223.38&14281.00&\textbf{2840.16}&679.63 &-114.00\\
MDG-c\_12\_n3000\_m500 &11366.00&\textbf{12001.40}&\textbf{12709.00}&3766.62&&\textbf{10604.00}&12103.03&13987.00&\textbf{2832.65}&782.51 &-762.00\\
MDG-c\_13\_n3000\_m500 &10942.00&\textbf{11832.40}&\textbf{12427.00}&3987.49&&\textbf{10743.00}&12228.58&14838.00&\textbf{2867.29}&949.06 &-199.00\\
MDG-c\_14\_n3000\_m500 &10903.00&\textbf{11455.20}&\textbf{12095.00}&3283.82&&\textbf{9941.00} &11643.90&15043.00&\textbf{2833.35}&1207.82&-962.00\\
MDG-c\_15\_n3000\_m500 &11051.00&\textbf{12311.90}&\textbf{13282.00}&3906.98&&\textbf{10870.00}&12365.85&14298.00&\textbf{2883.88}&757.94 &-181.00\\
MDG-c\_16\_n3000\_m600 &13934.00&\textbf{14732.10}&\textbf{15278.00}&4135.38&&\textbf{13910.00}&15801.65&18285.00&\textbf{2858.17}&1065.75&-24.00\\
MDG-c\_17\_n3000\_m600 &14086.00&\textbf{14882.70}&\textbf{16184.00}&4259.36&&\textbf{13676.00}&15284.10&17002.00&\textbf{2899.82}&712.02 &-410.00\\
MDG-c\_18\_n3000\_m600 &\textbf{13415.00}&\textbf{14515.20}&\textbf{15385.00}&4244.02&&14011.00&15547.08&16893.00&\textbf{2925.46}&589.07 &+596.00\\
MDG-c\_19\_n3000\_m600 &13850.00&\textbf{14821.90}&\textbf{15976.00}&4686.72&&\textbf{13538.00}&15526.85&16729.00&\textbf{2890.80}&753.32 &-312.00\\
MDG-c\_20\_n3000\_m600 &13532.00&14651.80&15396.00&4384.37&&\textbf{12415.00}&\textbf{13545.3}3&\textbf{14697.00}&\textbf{2879.54}&709.70 &-1117.00\\
\midrule[0.5pt]
Average value          &9862.40 &10662.29&\textbf{11448.30}&3033.69&&\textbf{9427.05} &\textbf{10544.52}&11932.70&\textbf{2791.86}&581.37 &-435.35\\
Wins                   & 1     & 9     & 11     & 10    && 19      & 11      & 9      & 10     & -     & -\\
\bottomrule[0.75pt]
\end{tabular}
\end{tiny}
\end{center}
\vskip -0.1in
\end{sidewaystable}

Finally, Table \ref{Tab:Comparison Summary} summarizes the performances of the proposed ILS\_MinDiff algorithm against the reference VNS\_MinDiff algorithm on all 190 benchmark instances. The significant differences are marked in bold. From this table, we can make the following observations:
\begin{itemize}
\item ILS\_MinDiff is highly competitive compared to the state-of-the-art results, and respectively wining 152.5, 154, 128 and 120.5 out of 190 instances in terms of the best solution value, the average solution value, the worst solution value and the average CPU time. 
\item  ILS\_MinDiff is significantly better than the current best-performing algorithm (VNS\_MinDiff) in terms of the best solution value on 4 out of 6 data sets (\textsl{SOM-b}, \textsl{MDG-a}, \textsl{MDG-b} and \textsl{MDG-c}). For \textsl{GKD-b} and \textsl{GKD-c}, ILS\_MinDiff respectively wins 31.5 and 14 instances, which are just slightly smaller than the corresponding critical values (i.e., $CV^{50}_{0.05} = 32$ and $CV^{20}_{0.05} = 15$).
\item Compared to the reference algorithm, ILS\_MinDiff significantly performs better in terms of the average solution value for all data sets except for \textsl{MDG-c}. Although the difference between ILS\_MinDiff and VNS\_MinDiff is not significant on \textsl{MDG-c}, ILS\_MinDiff still achieves a better average solution value for 11 out of 20 instances.
\item  ILS\_MinDiff performs significantly better than VNS\_MinDiff in terms of the worst solution value on \textsl{SOM-b}, \textsl{GKD-b}, \textsl{GKD-c} and \textsl{MDG-b}. For \textsl{MDG-a} and \textsl{MDG-c}, ILS\_MinDiff achieves a competitive performance, but the difference is not statistically significant.
\item In terms of computational efficiency, ILS\_MinDiff on average needs less time than VNS\_MinDif, but the significant difference is only confirmed on SOM-b and \textsl{GKD-c}. For the remaining four data sets, ILS\_MinDiff wins VNS\_MinDiff on most of instances, while the differences are not statistically significant.
\end{itemize}

\begin{table*}[!htbp]
\caption{A summary of win statistics between the proposed ILS\_MinDiff algorithm (left part of each column) and the reference VNS\_MinDiff algorithm (right part of each column) on all six data sets.}
\label{Tab:Comparison Summary}
\vskip 0.15in
\begin{center}
\begin{scriptsize}
\begin{tabular}{l|cccccc}
\toprule[0.75pt]
Indicator   & SOM-b & GKD-b & GKD-c & MDG-a & MDG-b & MDG-c\\
\midrule[0.5pt]
$f_{best}$  & $\textbf{18} \mid 2$    & $31.5 \mid 18.5$ & $14 \mid 6$ & $\textbf{33} \mid 7$      & $\textbf{37} \mid 3$  & $\textbf{19} \mid 1$\\
$f_{avg}$   & $\textbf{19} \mid 1$    & $\textbf{36} \mid 14$	   & $\textbf{20} \mid 0$	 & $\textbf{30} \mid 10$     & $\textbf{38} \mid 2$  & $11 \mid 9$\\
$f_{worst}$ & $\textbf{16} \mid 4$    & $\textbf{32} \mid 18$	   &	 $\textbf{20} \mid 0$	 & $21 \mid 19$     & $\textbf{30} \mid 10$ & $9 \mid 11$\\
$t_{avg}$   & $\textbf{15.5} \mid 4.5$& $26.5 \mid 23.5$ & $\textbf{19} \mid 1$	 & $23.5 \mid 16.5$	& $26 \mid 14$ & $10 \mid 10$\\
\bottomrule[0.75pt]
\end{tabular}
\end{scriptsize}
\end{center}
\vskip -0.1in
\end{table*}


\section{Experimental analysis}
\label{Sec:Experimental Analysis}

In this section, we study the impact of the parameters of ILS\_MinDiff on its performance: the search depth $nbr_{max}$, the weak perturbation strength $p_w$ and the strong perturbation coefficient $\alpha$. The analysis was based on a set of 9 instances selected from all six data sets, i.e., \textsl{SOM-b\_8\_n200\_m80}, \textsl{SOM-b\_20\_n500\_m200}, \textsl{GKD-b\_50\_n150\_m45}, \textsl{GKD-c\_20\_n500\_m50}, \textsl{MDG-a\_20\_n500\_m50}, \textsl{MDG-a\_40\_n2000\_m200}, \textsl{MDG-b\_20\_n500\_m50}, \textsl{MDG-b\_40\_n2000\_m200} and \textsl{MDG-c\_20\_n3000\_m600}. For each selected instance with $n < 2000$, we solved it 20 times, otherwise we solved it 10 times. 

\subsection{Effect of the search depth $nbr_{max}$}
\label{Effect of the search depth $nbr_{max}$}

\begin{figure}[!htbp]
\centering
\includegraphics[width=0.90\textwidth]{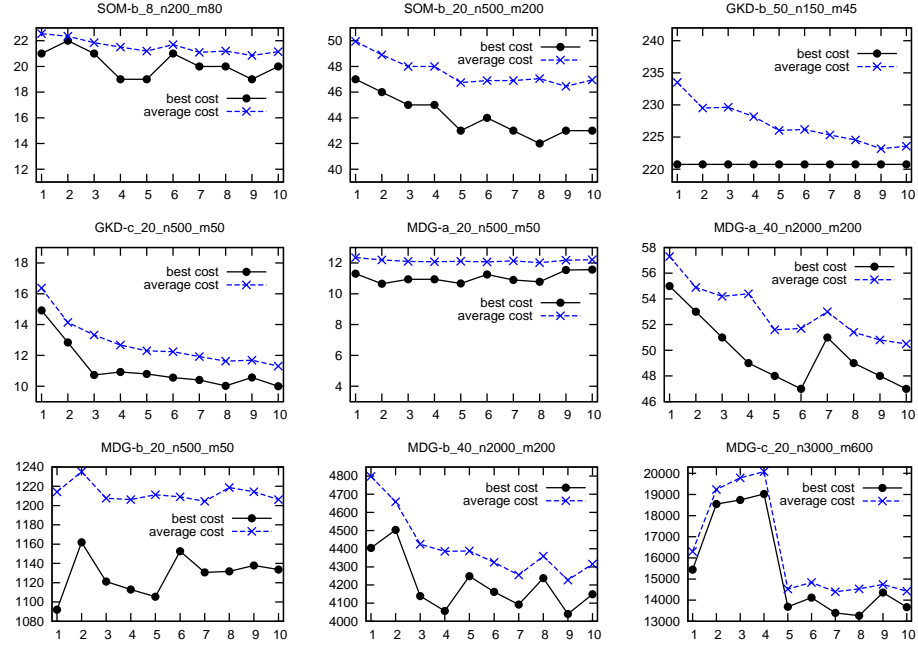}
\caption{Effect of the search depth $nbr_{max}$ on the performance of the ILS\_MinDiff algorithm}
\label{Fig:Search Depth}
\end{figure}

To investigate the effect of the search depth $nbr_{max}$ on the performance of ILS\_MinDiff, we first fixed the weak perturbation strength $p_w$ to 3 and the strong perturbation coefficient $\alpha$ to 1.0, i.e., $p_s = 1.0 \times  n/m$, and then varied $nbr_{max}$ from 1.0 to 10.0 with a step size of 1. Figure \ref{Fig:Search Depth} shows the behavior of the ILS\_MinDiff algorithm with  $nbr_{max}$ varying from 1.0 to 10.0, where the X-axis indicates the values of $nbr_{max}$ while Y-axis shows the best/average objective value.

Figure \ref{Fig:Search Depth} discloses that the average solution values of ILS\_MinDiff continuously decreases when the value of  $nbr_{max}$ increases from 1.0 to 10.0 for all tested instances except for \textsl{MDG-c\_20\_n3000\_m600}. For \textsl{MDG-c\_20\_n3000\_m600}, the curve shows a large variation and it also decreases when $nbr_{max}$ increases to 5. We also observe that ILS\_MinDiff obtains a superior best solution value when $nbr_{max}$ is 4 or 5 on all tested instances. This justifies the adopted setting ($nbr_{max}=5$) shown in Table \ref{Tab:Parameters Table}. 

\subsection{Effect of the weak perturbation strength $p_w$}

\begin{figure}[!htbp]
\centering
\includegraphics[width=0.90\textwidth]{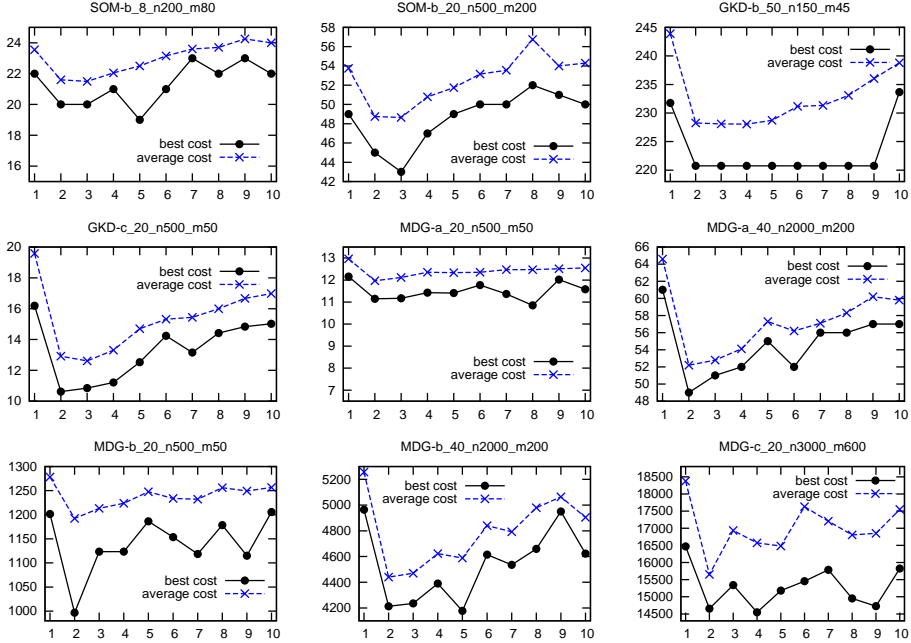}
\caption{Effect of the weak perturbation strength $p_w$ on the performance of the ILS\_MinDiff algorithm}
\label{Fig:Weak Perturbation Strength}
\end{figure}

To study the effect of the weak perturbation strength $p_w$ on the performance of the proposed algorithm, we first fixed the search depth $nbr_{max}$ to 5 according to the outcomes of Section \ref{Effect of the search depth $nbr_{max}$}, and the strong perturbation coefficient $\alpha$ to 1.0, i.e., $p_s = 1.0 \times n/m$. Then, we varied the value of $p_w$ from 1.0 to 10.0 with a step size of 1. Figure \ref{Fig:Weak Perturbation Strength} shows the performances of the ILS\_MinDiff with different values of $p_w$ varying from 1.0 to 10.0, where the X-axis and Y-axis respectively shows the weak perturbation strength $p_w$ and the best/average objective values.

Figure \ref{Fig:Weak Perturbation Strength} indicates that almost for all 9 tested instances, the best performance was attained when $p_w = 2$ or $p_w = 3$. A too large or small $p_w$ value gave a poor performance of the ILS\_MinDiff algorithm. This can be understood given that a small $p_w$ value cannot enable the search to jump out of the current local optimum, while a too large value of $p_w$ will have an effect similar to a random restart. This justifies the adopted setting for $p_w$  shown in Table \ref{Tab:Parameters Table}. 


\subsection{Effect of the strong perturbation coefficient $\alpha$}

\begin{figure}[!htbp]
\centering
\includegraphics[width=0.90\textwidth]{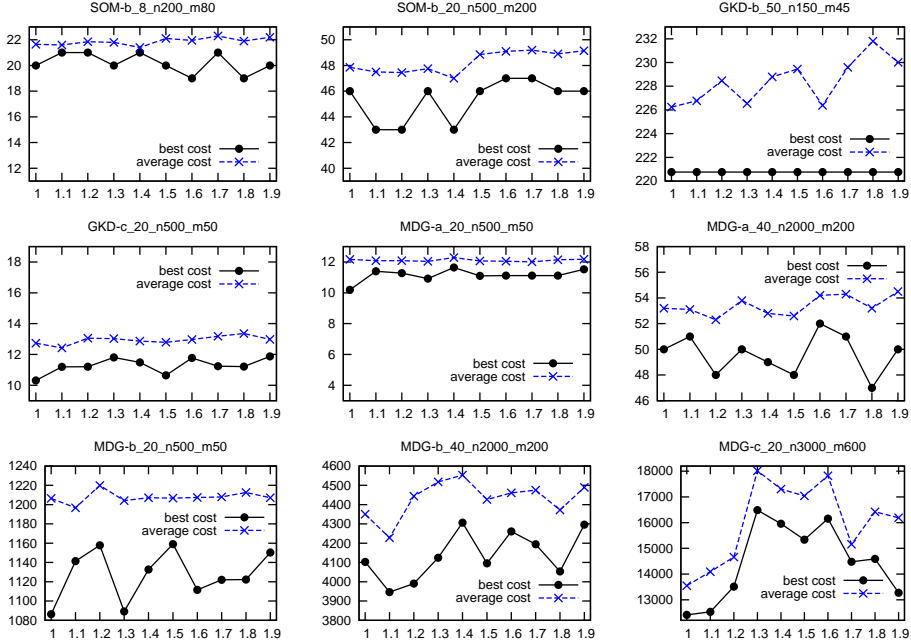}
\caption{Effect of the strong perturbation coefficient $\alpha$ on the performance of the ILS\_MinDiff algorithm}
\label{Fig:Strong Perturbation Coefficient}
\end{figure}

The ability of the ILS\_MinDiff algorithm to escape deep local optima depends on the strong perturbation strength $p_s = \alpha \times n/m$. To analyze the influence of the strong perturbation coefficient $\alpha$ on the performance of the algorithm, we fixed the search depth $nbr_{max}$ and the weak perturbation strength $p_w$ respectively according to the values determined from Figures \ref{Fig:Search Depth} and \ref{Fig:Weak Perturbation Strength}, and varied the strong perturbation coefficient $\alpha$ from 1.0 to 1.9 with a step size of 0.1. Figure \ref{Fig:Strong Perturbation Coefficient} displays the outcomes of this experiment,  where the X-axis and the Y-axis respectively denote the values of $\alpha$ and the best/average objective values.

As we can see from Figure \ref{Fig:Strong Perturbation Coefficient}, the average solution value is not really sensitive to $\alpha$ on almost all tested instances except for \textsl{MDG-c\_20\_n3000\_m600}. To achieve a better average solution value, we roughly set $\alpha = 1.0$ for all instances.

\section{Conclusions}
\label{Sec:Conclusions}

The minimum differential dispersion problem (\textsl{Min-Diff DP}) is a useful model in a variety of practical applications. However, finding high-quality solutions to large \textsl{Min-Diff DP} instances represents an imposing computational challenge. In this work, we have proposed a highly effective iterated local search algorithm for \textsl{Min-Diff DP} (denoted as ILS\_MinDiff), which adopts the general three-phase search framework. To ensure a suitable balance between intensification and diversification of the search process, ILS\_MinDiff runs sequentially and iteratively a fast descent-based neighborhood search phase to locate local optimal solutions, a local optima exploring phase to seek nearby better local optima, and a local optima escaping phase to escape deep attraction basin with strong perturbations. 

Extensive computational experiments on six data sets of 190 benchmark instances have demonstrated that despite its simplicity, the proposed algorithm competes very favorably with the state-of-the-art methods in the literature. In particular, ILS\_MinDiff is able to find new best results for 131 out of 190 instances (improved upper bounds) and match the best-known results for 42 out of the remaining 59 instances. These improved results can be used as new references for assessing other \textsl{Min-Diff DP} algorithms.




\end{document}